\documentclass[11pt,draftcls,onecolumn]{IEEEtran}
\usepackage{graphicx}
\usepackage{array}
\usepackage{amsmath}
\usepackage{amsfonts}
\usepackage{amsthm}
\usepackage{cite}
\usepackage{color}
\usepackage{caption}
\usepackage{subcaption}
\usepackage{cite}
\usepackage{multirow}
\usepackage{comment}
\usepackage{amssymb}
\usepackage{textcomp}

\newtheoremstyle{mythmstyle}
{3pt}
{3pt}
{\itshape}
{}
{\bfseries}
{:}
{.5em}
{}

\theoremstyle{mythmstyle}
\newtheorem{mydef} {Definition}
\newtheorem{mylemma} {Lemma}
\newtheorem{mypps} {Proposition}

\begin{document}
\title{Permutation Meets Parallel Compressed Sensing: How to Relax
Restricted Isometry Property for 2D Sparse Signals}
\author{Hao~Fang, Sergiy~A.~Vorobyov, Hai~Jiang, and Omid~Taheri%
\thanks{S.~A.~Vorobyov is the corresponding author. All authors are with the
  Department of Electrical and Computer Engineering, University of Alberta,
  Edmonton, AB, T6G~2V4, Canada; e-mail: \{{\tt hfang2, svorobyo, hai1,
  otaheri\}@ualberta.ca}.  S.~A.~Vorobyov is currently on leave and he is
	with Aalto Universuty, Finland. 

This work is supported in part by the Natural Science and Engineering Research Council (NSERC) of Canada. 

Some initial results on which this paper is built have been presented by the authors at Asilomar~2012, Pacific Grove, California, USA.}%
}
\maketitle

\begin{abstract}
Traditional compressed sensing considers sampling a 1D signal. For a multidimensional signal, if reshaped into a vector, the required size of the sensing matrix becomes dramatically large, which increases the storage and computational complexity significantly. To solve this problem, we propose to reshape the multidimensional signal into a 2D signal and sample the 2D signal using compressed sensing column by column with the same sensing matrix. It is referred to as {\it parallel compressed sensing}, and it has much lower storage and computational complexity. For a given reconstruction performance of parallel compressed sensing, if a so-called {\it acceptable permutation} is applied to the 2D signal, we show that the corresponding sensing matrix has a smaller required order of restricted isometry property condition, and thus, storage and computation requirements are further lowered. A zigzag-scan-based permutation, which is shown to be particularly useful for {signals satisfying a layer model}, is introduced and investigated. As an application of the parallel compressed sensing with the zigzag-scan-based permutation, a video compression scheme is presented. It is shown that the zigzag-scan-based permutation increases the peak signal-to-noise ratio of reconstructed images and video frames.
\end{abstract}

\begin{IEEEkeywords}
Compressed sensing, parallel processing, permutation, multidimensional signal processing.
\end{IEEEkeywords}

\section{Introduction}
\label{sec:intro}
Compressed sensing (CS) theory states that the information contained in an
$L$-length sparse signal ${\bf x}$ can be fully preserved with only $K \ll L$
measurements, which form a $K$-length vector ${\bf y}$ \cite{Donoho2006},
\cite{Candes2006}. 
This is done by the help of a $K \times L$ {\it sensing matrix} ${\bf A}$, i.e.,
${\bf y} = {\bf A x}$, where ${\bf A}$ satisfies the restricted isometry
property (RIP) of a certain order.
The signal ${\bf x}$ can be recovered from the $K$ measurements in ${\bf y}$ by
solving, for example, the following $\ell_1$-norm minimization problem 
\cite{Candes2005a}
\begin{equation}
\label{l1_min}
\min_{{\bf x}} ||{\bf x}||_1
\text{ s.t. } {\bf y} = {\bf A x}
\end{equation}
where $||\cdot||_1$ denotes the $\ell_1$-norm of a vector.

In addition, if signal ${\bf f}$ is not sparse itself, it can be represented
as a sparse signal in some orthonormal basis ${\bf \Psi}$, i.e., ${\bf x =
\Psi}^T{\bf f}$ is sparse signal. Here the superscript $T$ denotes the
transpose operation.
Then given the sensing matrix ${\bf A}$ for ${\bf x}$ and
the orthonormal basis ${\bf \Psi}$, the signal ${\bf f}$ can be measured using a
$K \times L$ {\it measurement matrix} ${\bf \Phi = A \Psi}^T$, i.e., ${\bf y = \Phi
f}$. 
It is equivalent to using ${\bf A}$ to sense ${\bf x}$ since ${\bf y = A
\Psi}^T {\bf f} = {\bf A x}$. Therefore, ${\bf x}$ and thus ${\bf f}$ can be
recovered from ${\bf y}$ as long as ${\bf A}$ satisfies the RIP of a certain
order. Consequently, two basic problems in CS are (i)~to find the basis ${\bf \Psi}$
{in which} the signal projection is sparse and (ii)~to construct sensing
matrix ${\bf A}$ and corresponding measurement matrix ${\bf \Phi}$.

Traditionally, CS is applied to sample 1D signals.
Recently, the research interest in applying CS to sample multidimensional
signals has increased significantly.
2D signals such as images and video frames are typical examples
of multidimensional signals.
A branch of CS theory, named {\it compressive
imaging}, is introduced in \cite{Wakin2006}, where the {\it single-pixel
camera} is proposed.
The single-pixel camera acquires a group of measurements
of an image using different patterns of the digital micromirror device (DMD)
array, without collecting the pixels.
Mathematically, each pattern of the DMD
array plays the role of a row in the measurement matrix ${\bf \Phi}$, and the
image is viewed as a vector.
Consequently, the size of the DMD array is the same
as the expected number of pixels in the image. The research is extended to
sample color images by combining the Bayer color filter and the DMD array
\cite{Nagesh2009}.
Furthermore, for reconstruction, the architecture proposed in
\cite{Nagesh2009} employs joint sparsity models to exploit the correlation among
different color channels.
However, as the expected number of pixels in the image increases, the number of
columns and the required number of rows in the measurement matrix ${\bf \Phi}$
also increase. 
In other words, both the size and the required number of
patterns of the DMD array increase.
Therefore, the implementation cost and the complexity of the encoder increase
significantly as well.
For example, the storage of ${\bf \Phi}$ and computational complexity for
acquiring measurements are unwieldy for any normal size images.
A general framework for sampling multidimensional signals, named {\it
Kronecker CS}, is proposed in \cite{Duarte2012}.
In Kronecker CS, a multidimensional signal is vectorized and then sampled using
a measurement matrix which is the Kronecker product of several smaller-sized
measurement matrices that correspond to the measurement processes for different
portions of the multidimensional signal\footnote{For example, for a 2D signal,
the smaller-sized measurement matrices can correspond to the measurement
processes for rows and for columns of the 2D signal, respectively.}.
After finding such Kronecker product, the resulting measurement matrix
clearly has a very large size.  
Thus, the problem related to the storage of ${\bf \Phi}$ and computational
complexity for acquiring measurements arises as well in the Kronecker CS. 

To address the above problem, instead of storing the Kronecker product of 
several measurement matrices, all portions of the multidimensional signal 
can be sampled sequentially using corresponding smaller-sized measurement 
matrices. Then the encoder needs to store only the smaller-sized measurement 
matrices. Using this approach, a separable sensing operator is designed for 
compressive imaging in \cite{Rivenson2009}, where an imaging operator (the 
measurement matrix for the whole image) can be separated into two dimensions. 
The separable sensing operator design significantly reduces the complexity
of implementation, storage, and usage of the imaging operator. Another 
solution to the problem of storage and computational complexity is the block 
CS of \cite{Gan2007}. The idea is to divide a 2D signal into smaller blocks 
and sample individual vectorized blocks, whereas all blocks need to be
reconstructed as a whole. The block CS of \cite{Gan2007} uses a block-diagonal 
measurement matrix for sampling the vectorized signal ${\bf x}$.
As a result, the block CS can reduce the storage and computational complexity 
at the encoder side. Some improved reconstruction algorithms for the block CS 
scheme are presented in \cite{Mun2010}. They help to further reduce the 
required number of rows in the measurement matrix ${\bf \Phi}$ for a given 
reconstruction error requirement. Based on the block CS architecture, a fast 
sampling operator is proposed in \cite{Gan2008} using the block Hadamard 
ensemble, which can be easily implemented in the optical domain. 

The use of fast algorithms for computing measurements by taking advantage of 
the measurement matrix's structure is another way to address the problem of 
storage and computational complexity at the encoder side. For example, in
\cite{Candes2006b} and \cite{Candes2005}, a scrambled Fourier ensemble is 
used as the sensing matrix ${\bf A}$ and the wavelet basis is used as the 
orthonormal basis ${\bf \Psi}$ in which the image projection is sparse.
Thus, the sampling process can be implemented efficiently by first 
transforming a 2D image into the wavelet domain and then applying ${\bf A}$ 
to the wavelet coefficients by means of the fast Fourier transform. The 
sparsity structure of the multidimensional signal can be employed as well
to reduce the storage and computational complexity at the encoder side.
It is proposed in \cite{Han2008} to decompose the wavelet coefficients of a 
2D signal into sparse components and dense components, and apply CS only 
to sparse components using a smaller-sized sensing matrix. In \cite{Wu2011} 
and \cite{Kim2012}, the statistical structure of the wavelet coefficients 
of a 2D image is considered for CS reconstruction of the image by using 
a scale mixture model. It is shown that fewer measurements are required 
in order to achieve a given reconstruction error performance. In addition, 
the scheme in \cite{Wu2011} suggests to rearrange the wavelet coefficients 
into a new 2D matrix and sample each column of the matrix using the same 
sensing matrix ${\bf A}$ of a smaller size.

All above research works focus on the encoder side, aiming at reducing 
the implementation cost and the storage and computational complexity of 
the encoder. Joint reconstruction is employed in these schemes, and thus, 
the complexity at the decoder side is still high.

Taking into account the implementation cost and the storage and 
computational complexity of the decoder, a block-based CS architecture 
can be employed in video compression where all blocks can be sampled and 
reconstructed independently. In \cite{Stankovic2008b} and \cite{Stankovic2008}, 
it is proposed to apply CS only to sparse blocks found by a block 
classification scheme that considers the difference of sparsity levels 
among different blocks in an image or a video frame. Another block 
classification scheme based on inter-frame correlation is proposed in 
\cite{Liu2011}.

In this paper, a parallel CS scheme is developed. A multidimensional 
sparse signal is considered, i.e., the signal is sparse in the identity basis. 
The multidimensional signal is first rearranged into a 2D matrix,
and then sampled column-by-column via CS using the same sensing matrix.
In this way, the required size of the sensing matrix can be reduced
significantly compared to the scheme that samples the vectorized signal.
Furthermore, both sampling and reconstruction can be conducted for individual
columns in parallel. Note that several works use a similar column-by-column 
sampling setting at the encoder side, e.g., the aforementioned scheme of 
\cite{Wu2011}. The focus of \cite{Wu2011} is on studying the scale mixture 
models used in CS for image reconstruction. The signal considered in 
\cite{Wu2011} is the matrix of wavelet coefficients. Another example is the 
multiple measurement vectors (MMV) model of \cite{Cotter2005}, which considers 
a group of signals that share the same sparsity profile. In the MMV model, a 
group of signals are sampled using the same dictionary, which is analogous to 
the sensing matrix in CS, while the group of signals can be viewed as a 
virtual 2D signal. Joint reconstruction is then used for the MMV model.
Compared to the aforementioned works \cite{Wu2011} and \cite{Cotter2005}, 
which consider some specific sparse signals and require joint reconstruction 
at the decoder side, we address a more general setting at the encoder side and 
develop a parallel reconstruction at the decoder side. Moreover, we derive 
some analytical results related to the parallel CS scheme. Although joint 
reconstruction of multiple vectors, e.g., reconstruction of multiple vectors 
via sum-of-norm minimization, can bring some benefits \cite{Chen2006}, it is 
shown that the uniform-recovery rate in the sum-of-norm minimization case 
cannot exceed that for the case of individual reconstruction of each vector 
\cite{Berg2010}. Besides, there are problems that cannot be solved by the 
joint recovery via sum-of-norm minimization, but can be solved by 
individually and independently reconstructing each vector \cite{Berg2010}.

In the parallel CS architecture proposed in this paper, a 2D signal may 
be permuted before it is sampled. It is because the permulation may 
provide benefits, for example, in computation and storage. Permutations 
are studyed in several papers related to CS, though the goals of permutations 
in the exisitng literature are very different from our goal here. In 
\cite{Omid2011}, a segmented CS architecture is proposed and it is shown that
a similar improvement to that obtained by increasing the size of the measurement
matrix can be achieved by using a virtual extended measurement matrix obtained
by permuting the existing rows of the initial measurement matrix. In 
\cite{Duarte2012a}, it is shown that if nonzero entries of a sparse signal 
are clustered, the deterministic Delsarte-Goethals frame used as sensing 
matrix does not work. Thus, it is poroposed to apply permutations to the 
signal in order to avoid clustered nonzero entries. In our paper, the goal 
for applying permutations is different. Specifically, the parallel CS 
architecture considers sensing matrices that satisfy the RIP, and permutation 
is applied to 2D-reshaped signal aiming at ensuring that all columns of such 
signal have similar sparsity levels. We show that if a so-called acceptable 
permutation is conducted before sampling, the sensing matrix needs to satisfy 
the RIP of a smaller order than the sensing matrix of the parallel CS without 
any permutation. Thus, the storage and computational complexity can be 
further reduced. In our paper, a group-scan-based permutation is introduced for 
2D signals which can be divided into a number of groups with elements in each 
group having the same probability to be large in magnitude. As a special case 
of such group-scan-based permutation, a zigzag-scan-based permutation is 
introduced and investigated for 2D signals satisfying a newly introduced layer 
model. A video compression scheme based on the proposed parallel CS with the
zigzag-scan-based permutation is also developed and investigated. It improves
the peak signal-to-noise ratio (PSNR) of reconstructed frames compared {to} 
parallel CS without permutation. This demonstrates the effectiveness of the
zigzag-scan-based permutation in image compression.

In summary, this paper makes four main contributions. First, we propose a
parallel CS scheme, which reduces the required size of sensing matrix and
can be conducted at both the encoding and decoding sides in a parallel
(column-by-column) manner. Second, we investigate properties of permutations
when applied to parallel CS. Third, we introduce a group-scan-based
permutation and a zigzag-scan-based permutation and as an example we show
that the zigzag-scan-based permutation is an acceptable permutation with a 
large probability for 2D signals satisfying a newly introduced layer model. 
Finally, an application of the proposed parallel CS with the zigzag-scan-based 
permutation to video compression in wireless multimedia sensor networks is 
discussed. Some very preliminary results have been reported in 
\cite{Fang2012Asilomar}.

The remainder of the paper is organized as follows. Section~\ref{pcs} introduces
the parallel CS scheme. Permutations are discussed in Section~\ref{perm}.
Section~\ref{cvs} describes the video compression scheme that employs parallel
CS with the zigzag-scan-based permutation in application to wireless multimedia 
sensor networks. Simulation results are given in Section~\ref{sim}. Finally, 
conclusions are given in Section~\ref{conclusion}. The software needed to
generate the numerical results can be obtained from
http://www.ualberta.ca/\texttildelow hfang2/pub/2013TSP.zip.

\section{Parallel CS}
\label{pcs}
Given any multidimensional sparse signal, we can rearrange it into a 2D matrix
${\bf X} \in \mathbb{R}^{M \times N}$. A multidimensional signal and the
corresponding 2D matrix ${\bf X}$ are called $s$-sparse or have sparsity level
$s$ if ${\bf X}$ has only $s$ nonzero entries. The sparsity level of ${\bf X}$
can be denoted as a sparsity vector ${\bf s} = [s_1, s_2, \cdots, s_N]$, where
$s_j$ is the sparsity level of the $j$-th column {of ${\bf X}$}. In other words,
the $j$-th column of ${\bf X}$ has only $s_j$ nonzero entries. Apparently,
$||{\bf s}||_1 = s$.

In terms of the 2D signal ${\bf X}$, the proposed parallel CS consists of
sampling each column of ${\bf X}$ by the same sensing matrix ${\bf A}$ and
reconstructing these columns individually and in parallel by using any 1D CS
reconstruction algorithm. In this paper, for presentation simplicity, we
consider 2D signals, i.e., rearrangement of a multidimensional signal into a
2D matrix is done in advance.

\subsection{Theoretical Results on CS for 1D Signals}
Most practical signals are not strictly sparse, but rather regarded as
\emph{compressible}, i.e., they have only a few large\footnote{In this paper,
when we say a value is large or small, it means the magnitude of the value is
large or small.} elements. Here we use a 1D signal ${\bf x}$ as an example. The
signal ${\bf x}$ can be approximated using its \emph{best $s$-term
approximation} {denoted as} ${\bf x}^s$, which is an $s$-sparse signal generated
by keeping the $s$ largest entries in ${\bf x}$ and changing the remaining
entries to zeros. The best $s$-term approximation is regarded as an optimal
approximation using only $s$ elements. However, such approximation requires
knowledge about the values and locations of all elements in ${\bf x}$.

On the other hand, when CS is applied to ${\bf x}$, it is known that, if the
sensing matrix ${\bf A}$ obeys the RIP of order $s$, the reconstruction via
solving (\ref{l1_min}) is nearly as good as that using the best $s$-term
approximation, as shown in the following Lemma~\ref{lemma:RIP}
\cite{Candes2005a}, \cite{Candes2008}.

\begin{mydef}{\cite{Candes2005a}}
For every integer $s = 1, 2, \dots,$ the $s$-restricted isometry constant
$\delta_s$ of a given matrix ${\bf A}$ is defined as the smallest quantity such
that the inequality
\[
(1-\delta_s)||{\bf z}||_2^2 \leq ||{\bf Az}||_2^2 \leq (1+\delta_s)||{\bf z}||_2^2
\]
holds for all sparse signals ${\bf z}$ with no more than $s$ nonzero entries,
where $||\cdot||_2$ denotes the $\ell_2$-norm of a vector.
\end{mydef}

\begin{mylemma}{\cite{Candes2008}}
\label{lemma:RIP}
Assume that $\delta_{2s} < \sqrt{2}-1$ for a sensing matrix ${\bf A}$. Then for
a signal ${\bf x}$, the solution ${\bf x}^*$ to (\ref{l1_min}) obeys
\[
||{\bf x}^* - {\bf x}||_1 \leq G \cdot ||{\bf x - x}^s||_1 \quad \text{and}
\]
\begin{equation}
\label{eq:RIP_error}
||{\bf x}^* - {\bf x}||_2 \leq G' \cdot ||{\bf x - x}^s||_1 / \sqrt{s}
\end{equation}
for some constants $G$ and $G'$.
\end{mylemma}

In this paper, if the RIP condition holds for ${\bf A}$ with $\delta_{2s}~<~
\sqrt{2} - 1$, the matrix ${\bf A}$ is regarded as obeying the RIP of order $s$.
Therefore, according to Lemma~\ref{lemma:RIP}, an $s$-sparse signal
can be exactly reconstructed via solving (\ref{l1_min}) if the sensing
matrix A obeys the RIP of order $s$. For a compressible signal ${\bf x}$, if a
sensing matrix ${\bf A}$ obeying the RIP of order $s$ is used to sample ${\bf
x}$, the reconstruction via solving (\ref{l1_min}) has an error bounded by the
$\ell_1$-norm of the approximation error when ${\bf x}^s$ is used to approximate
${\bf x}$. Note that, for reconstruction via solving (\ref{l1_min}), we do not
need knowledge about the values and locations of all elements in ${\bf x}$,
while such knowledge is needed if ${\bf x}^s$ is used to approximate ${\bf x}$.

\subsection{New Theoretical Results on Parallel CS for 2D Signals}
Based on Lemma \ref{lemma:RIP}, the following lemma gives a sufficient condition
for exact reconstruction of a 2D $s$-sparse signal using parallel CS.

\begin{mylemma}
\label{lemma:RIP_2D_sparse}
Consider a 2D $s$-sparse signal ${\bf X}$ with sparsity vector ${\bf s}$, if the
RIP of order $||{\bf s}||_\infty$ holds for the sensing matrix ${\bf A}$, i.e.,
$\delta_{2||{\bf s}||_\infty} < \sqrt{2} - 1$, where $||\cdot||_\infty$ stands
for the {Chebyshev norm of a vector\footnote{The Chebyshev norm of a vector is
equal to the largest magnitude of the elements in the vector.},} then ${\bf X}$
can be exactly reconstructed using parallel CS scheme.
\end{mylemma}
\begin{IEEEproof}
The proof follows the same steps as the proof for the following Lemma
\ref{lemma:RIP_2D_compressible}.
\end{IEEEproof}

For a 2D compressible signal ${\bf X}$, the following lemma gives a sufficient
condition that the reconstruction error of the parallel CS is bounded by the
$\ell_1$-norm of the approximation error when the best $s$-term approximation of
${\bf X}$, {denoted as} ${\bf X}^s$, is used to approximate ${\bf X}$.
\begin{mylemma}
\label{lemma:RIP_2D_compressible}
Let ${\bf X}^s \in \mathbb{R}^{M \times N}$, which has a sparsity vector ${\bf
s} = [s_1, s_2, \cdots, s_N]$, be the best $s$-term approximation of ${\bf
X}~\in~\mathbb{R}^{M \times N}$. If the sensing matrix ${\bf A}$ obeys the RIP
of order $||{\bf s}||_\infty$, i.e., $\delta_{2||{\bf s}||_\infty} < \sqrt{2} -
1$, then the signal $\hat{{\bf X}}$ reconstructed using parallel CS scheme obeys
\[
||\hat{{\bf X}} - {\bf X}||_1
\leq G \cdot ||{\bf X} - {\bf X}^s||_1 \quad \text{and}
\]
\begin{equation}
||\hat{{\bf X}} - {\bf X}||_2
\leq G' \cdot ||{\bf X} - {\bf X}^s||_1
\label{eq:RIP_2D_error}
\end{equation}
where $G$ and $G'$ are finite constants.
\end{mylemma}
\begin{IEEEproof}
Since for all $1 \leq j \leq N$, $||{\bf s}||_\infty \geq s_j$, {and} according
to the definition of $s$-restricted isometry constant, we have $\delta_{2s_j}
\leq \delta_{2||{\bf s}||_\infty} < \sqrt{2} - 1$. Then, according to
(\ref{eq:RIP_error}), we {obtain that}
\begin{equation*}
||\hat{{\bf x}}_j - {\bf x}_j||_1
\leq G_j \cdot ||{\bf x}_j - {\bf x}^s_j||_1
\end{equation*}
and
\begin{equation*}
||\hat{{\bf x}}_j - {\bf x}_j||_2
\leq G'_j \cdot ||{\bf x}_j - {\bf x}^s_j||_1 \cdot s_j^{-1/2}
\end{equation*}
where ${{\bf x}}_j$, $\hat{{\bf x}}_j$, and ${\bf x}^s_j$ denote the $j$-th column of ${\bf
X}$, $\hat{\bf
X}$, and ${\bf X}^s$, respectively, and $G_j$ and $G_j'$ are
finite constants. Therefore, by choosing {$G = \max_j\{G_j\}$ and $G' =
\max_j\{G'_j\}$}, we obtain that
%\begin{flalign*}
\begin{equation*}
\begin{array}{ll}
\qquad &||\hat{{\bf X}} - {\bf X}||_1
= \sum_{i=1}^N ||\hat{{\bf x}}_j - {\bf x}_j||_1 \\
&\leq G \cdot \sum_{j=1}^N ||{\bf x}_j - {\bf x}^s_j||_1
= G \cdot ||{\bf X} - {\bf X}^s||_1
\end{array}
\end{equation*}
%\end{flalign*}
and
%\begin{flalign*}
\begin{equation*}
  \begin{array}{ll}
\qquad &||\hat{{\bf X}} - {\bf X}||_2
= \sqrt{\sum_{j=1}^N ||\hat{{\bf x}}_j - {\bf x}_j||_2^2} \\
&\leq \sqrt{G'^2 \cdot \sum_{i=1}^N ||{\bf x}_j - {\bf x}^s_j||_1^2}
= G' \cdot \sqrt{\sum_{j=1}^N ||{\bf x}_j - {\bf x}^s_j ||_1^2} \\
&\leq G' \cdot \sum_{j=1}^N ||{\bf x}_j - {\bf x}^s_j||_1
= G' \cdot ||{\bf X} - {\bf X}^s||_1.
\end{array}
\end{equation*}
%\end{flalign*}

This completes the proof.
\end{IEEEproof}

To sum up, for parallel CS, the RIP condition for the sensing matrix ${\bf A}$
for a given reconstruction quality is related to $||{\bf s}||_\infty$. In
Subsection~\ref{ssec:dis_perm}, it will be shown that the RIP condition can be
relaxed by performing a so-called acceptable permutation before using the
parallel CS.

\section{Permutation}
\label{perm}
When parallel CS is applied to a 2D compressible signal\footnote{Without loss of
generality, compressible signals are considered in the remainder of the paper,
since a sparse signal can be regarded as a special case of a compressible
signal.} ${\bf X}$, {the} difference of sparsity levels among columns of ${\bf
X}^s$ (which has sparsity vector ${\bf s}$) is not considered, and thus, the
{`worst-case'} sparsity level of the columns of ${\bf X}^s$, i.e., $||{\bf
s}||_\infty$, needs to be taken into account when designing the sensing matrix.
In this section, permutation is introduced {such that} by permuting\footnote{In
this paper, when we say ``permute", it means exchanging entries in a 2D matrix,
while not changing the dimension of the matrix.} entries {of ${\bf X}$} all
columns of the best $s$-term approximation of the newly formed 2D signal {would}
share similar sparsity {levels}. 

Let P$(\cdot)$ be a permutation operator which maps a matrix into another matrix
by permuting its elements and P$^{-1}(\cdot)$ be the corresponding inverse
permutation operator. Then ${\bf X}^\dag = \text{P}({\bf X})$ and ${\bf X} =
\text{P}^{-1}({\bf X}^\dag)$ where ${\bf X}^\dag \in \mathbb{R}^{M \times N}$ is
a permuted 2D signal.

With permutation before sampling, the parallel sampling process {can be}
described as follows
\begin{equation}
\label{eq:parallel_enc}
{\bf y}_j = {\bf A x}_j^\dag
\end{equation}
where ${\bf x}_j^\dag$ is the $j$-th column of ${\bf X}^\dag$, and ${\bf y}_j$
is the measurement vector of ${\bf x}_j^\dag$.
We can rewrite (\ref{eq:parallel_enc}) in the matrix form as
\begin{equation}
{\bf Y} = {\bf A X}^\dag
= {\bf A} \text{P}({\bf X})
\end{equation}
where ${\bf Y} = [{\bf y}_1, {\bf y}_2, \cdots, {\bf y}_N]$.

For signal reconstruction, all columns of ${\bf X}^\dag$ can be reconstructed
in parallel by any existing CS reconstruction algorithm. Let 
$\hat{\bf X}^\dag$ be the reconstructed permuted signal. Then we can apply 
inverse permutation to $\hat{\bf X}^\dag$ to obtain the reconstructed 2D signal 
$\hat{{\bf X}}$, that is,
\begin{equation}
\hat{{\bf X}} = \text{P}^{-1}(\hat{{\bf X}}^\dag).
\end{equation}

\subsection{Discussion about Permutation}
\label{ssec:dis_perm}
For any multidimensional signal, the permutation can be either applied after or
included in the process of rearranging the multidimensional signal into a 2D
matrix. The block-based CS employed in \cite{Gan2007}, \cite{Mun2010} and
\cite{Stankovic2008} is a special case of the parallel CS, which can be
interpreted as making each vectorized block as a column of a new 2D
signal.
Furthermore, the problem of difference of sparsity levels among blocks is
addressed in \cite{Stankovic2008} by employing a classification scheme to
identify sparse blocks and dense blocks and then applying CS only to the
sparse blocks.
In our work, permutation is applied to
a 2D compressible signal ${\bf X}$ or integrated into the process of
rearrangement of a multidimensional signal to a 2D compressible signal ${\bf X}$
such that all columns of ${\bf X}^{\dag s}$ (the best $s$-term approximation of
the resulted 2D signal ${\bf X}^\dag$) are sparse. Thus, the classification step of
\cite{Stankovic2008} is avoided.

Consider a compressible 2D signal ${\bf X}$ and its best $s$-term approximation
${\bf X}^s$ with sparsity vector ${\bf s}$ (then we have $||{\bf s}||_1~=~s$).
If the sensing matrix ${\bf A} \in \mathbb{R}^{K \times M}$ is constructed from
Gaussian ensembles with
\begin{equation}\label{e:Kbound}
K \geq C \cdot ||{\bf s}||_\infty \log{(M/||{\bf s}||_\infty)}
\end{equation}
for some constant $C$, then it will satisfy the RIP of order $||{\bf
s}||_\infty$ \cite{Candes2005a}. Then according to
Lemma~\ref{lemma:RIP_2D_compressible}, the signal $\hat{\bf X}$ reconstructed
using parallel CS obeys (\ref{eq:RIP_2D_error}).

\begin{mydef}
  \label{def:accep_perm}
For a 2D compressible signal ${\bf X} \in \mathbb{R}^{M \times N}$ and its best
$s$-term approximation ${\bf X}^s$ with sparsity vector ${\bf s}$, a permutation
${\text{P}}(\cdot)$ is called {\it acceptable} for ${\bf X}$ if the Chebyshev
norm of the sparsity vector of the best $s$-term approximation of the signal
P$({\bf X})$ is smaller than $||{\bf s}||_\infty$.
\end{mydef}

When permutation is applied before parallel CS, the signal after permutation is
${\bf X}^\dag$, and the best $s$-term approximation of ${\bf X}^\dag$ is
{denoted as} ${\bf X}^{\dag s}$ with sparsity vector ${\bf s}^\dag$ (then we
have $||{\bf s}^\dag||_1=s$).
Consider that $M \gg ||{\bf s}||_\infty$ and $M \gg ||{\bf s}^\dag||_\infty$,
i.e., ${\bf X}^s$ and ${\bf X}^{\dag s}$ are sparse enough. If $||{\bf
s}^\dag||_\infty < ||{\bf s}||_\infty$, it can be seen that for parallel CS with
{an acceptable} permutation, the lower bound of $K$ in (\ref{e:Kbound}) is
smaller than that in parallel CS without {an acceptable permutation}. In other
words, for the sufficient condition in Lemmas
\ref{lemma:RIP_2D_sparse}~and~\ref{lemma:RIP_2D_compressible}, {the} condition
``${\bf A}$ obeys the RIP of order $||{\bf s}^\dag||_\infty$" for parallel CS
with {an acceptable permutation is weaker than {the} condition ``${\bf A}$ obeys
the RIP of order $||{\bf s}||_\infty$" for parallel CS without an acceptable
permutation.}
%if $||{\bf s}^\dag||_\infty < ||{\bf s}||_\infty$. 
To sum up, the RIP condition for a given reconstruction quality is weaker after 
permutation if $||{\bf s}^\dag||_\infty$ is smaller than $||{\bf s}||_\infty$.

Since $||{\bf s}^\dag||_1=||{\bf s}||_1=s$, it is desired that, after an
acceptable permutation, the $s$ nonzero elements in the best $s$-term
approximation of the permuted 2D signal are evenly distributed among the
columns, which leads to minimum $||{\bf s}^\dag||_\infty$. Such a permutation is
an optimal permutation defined below.

% \subsection{Optimal Permutation}
% \label{opt_acc_perm}
\begin{mydef}
For a 2D compressible signal ${\bf X} \in \mathbb{R}^{M \times N}$ and its best $s$-term approximation ${\bf X}^s$, if after a permutation,
the best $s$-term approximation ${\bf X}^{\dag s}$ of the resulted 2D signal ${\bf X}^\dag$ has sparsity vector ${\bf s}^*$ satisfying $\max_i \{s^*_i\} - \min_i \{s^*_i\} \leq 1$, where $s^*_i$ denotes the $i$-th
entry of ${\bf s}^*$,
then ${\bf s}^*$ is called an optimal sparsity vector of ${\bf X}^s$, and the
corresponding permutation is call an optimal permutation of ${\bf X}$.
\end{mydef}

%\begin{mydef}
%For a 2D compressible signal ${\bf X} \in \mathbb{R}^{M \times N}$, its $(MN)!$
%possible permutations result in $(MN)!$ best $s$-term approximations.
%Among the $(MN)!$ sparsity vectors of the $(MN)!$
%best $s$-term approximations, 
%if a sparsity vector ${\bf s}^*$ satisfies $||{\bf s}^*||_1 = s$ and 
%$\max_i \{s^*_i\} - \min_i \{s^*_i\} \leq 1$, where $s^*_i$ denotes the $i$-th
%entry of ${\bf s}^*$, 
%such sparsity vector is called an optimal sparsity vector of ${\bf X}^s$, and the
%corresponding permutation is call an optimal permutation of ${\bf X}$.
%\end{mydef}

\begin{mylemma}
\label{lemma:opt_perm_necc}
For a 2D compressible signal ${\bf X} \in \mathbb{R}^{M \times N}$ and its best
$s$-term approximation ${\bf X}^s$, there exists at least one optimal sparsity
vector ${\bf s}^*$ of ${\bf X}^s$.
\end{mylemma}
\begin{IEEEproof}
  Obviously, $||{\bf s}^*||_1 = s$.
  If $\lceil s/N \rceil = \lfloor s/N \rfloor = s/N$ (where $\lceil \cdot
  \rceil$ denotes the ceiling function and $\lfloor \cdot \rfloor$ denotes the
  floor function), we can immediately find an
  optimal sparsity vector ${\bf s}^*$ whose entries are all $s/N$.
  If $\lceil s/N \rceil \neq \lfloor s/N \rfloor$, we consider a permutation on
  ${\bf X}$ such that: for the best $s$-term approximation of the resulted 2D
  signal, there are $\lceil s / N \rceil$ nonzero elements in each of the first
  $s - N \lfloor s / N \rfloor$ columns, and the remaining nonzero elements are
  evenly distributed among the remaining columns.
  Then each of the last $N \lceil s/N \rceil - s$ columns has $\lfloor s/N
  \rfloor$ nonzero entries. 
  Therefore, the sparsity vector of the best $s$-term approximation of the permuted 2D signal is an optimal sparsity vector.
  This completes the proof.
 \end{IEEEproof}

From the proof of Lemma~\ref{lemma:opt_perm_necc}, it follows that optimal sparsity vector and optimal permutation may not be
unique, and the
Chebyshev norm of an optimal sparsity vector of ${\bf X}^s$ is equal to $\lceil
s/N \rceil$.

\begin{comment}
The following lemma gives a sufficient condition for a sparsity vector to be 
an optimal sparsity vector.
\begin{mylemma}
\label{lemma:opt_perm_suff}
For the best $s$-term approximation of a 2D compressible signal ${\bf X} \in
\mathbb{R}^{M \times N}$, {denoted as} ${\bf X}^s$, if a sparsity vector ${\bf
s}=[s_1,s_2,\cdots,s_N]$ satisfies $||{\bf s}||_1 = s$ and $\max_j\{s_j\} -
\min_j\{s_j\}~\leq~1$, then ${\bf s}$ is an optimal sparsity vector of ${\bf
X}^s$.
\end{mylemma}
\begin{IEEEproof}
We denote $l$ as the number of elements in ${\bf s}$ that are equal to
$\max_j\{s_j\}$. Then all other $(N-l)$ elements in ${\bf s}$ are equal to
$\min_j(s_j)$, since {$\max_j\{s_j\} - \min_j\{s_j\}~\leq~1$.} So we have

\begin{align*}
s&=||{\bf s}||_1 =l\max_j\{s_j\} + (N-l)\min_j\{s_j\} \\
&=\!N\max_j\{s_j\}-(N-l)\big(\max_j\{s_j\}-\min_j\{s_j\}\big) \\ 
&\ge N\max_j\{s_j\}-(N-l).
\end{align*}
Thus, for any optimal sparsity vector of ${\bf X}^s$, denoted as ${\bf s}^*$,
using Lemma~\ref{lemma:opt_sparsity}, we have
\begin{align*}
||{\bf s}^*||_\infty = \Big\lceil \frac{s}{N} \Big\rceil 
&\ge \Big\lceil \frac{N\max_j\{s_j\}-(N-l)}{N} \Big\rceil \\ 
&= \max_j\{s_j\}=||{\bf s}||_\infty.
\end{align*}

Then apparently ${\bf s}$ is an optimal sparsity vector of ${\bf X}^s$. {This 
completes the proof.}
\end{IEEEproof}
\end{comment}

In most scenarios, finding an optimal permutation may not be practical. An
acceptable permutation defined in Definition~\ref{def:accep_perm} can be used
instead.

\subsection{Group-scan-based Permutation and Zigzag-scan-based Permutation}
\label{zigzag_perm}
The following observation is of interest. 
For a 2D compressible signal ${\bf X} \in \mathbb{R}^{M \times N}$, consider a
permuted signal ${\bf X}^\dag$ and its best $s$-term approximation ${\bf
X}^{\dag s}$. For any {$1 \leq i \leq N$},
if all elements in the $i$-th row of ${\bf X}^{\dag s}$ share the same
probability to be nonzero, {denoted as} $p_i$, then all columns of ${\bf
X}^{\dag s}$ have the same expected sparsity level, given as $\sum_{i=1}^M p_i$.

For example, when $M=N=4$, if after a permutation, the elements in the 1st, 2nd,
3rd and 4th rows of ${\bf X}^{\dag s}$ have {respectively} probabilities $p_1 =
0.9$, $p_2 = 0.3$, $p_3 = 0.2$ and $p_4 = 0.1$ to be nonzero, then for the
sparsity vector of ${\bf X}^{\dag s }$, denoted as ${\bf
s}^\dag=[s_1^\dag,s_2^\dag,s_3^\dag,s_4^\dag]$, we have
\[
{\rm E}\left\{\max_j\{s_j^\dag\} - \min_j\{s_j^\dag\}\right\} = 1.3881
\]
and
\begin{equation}
\label{eq:pr_opt_perm}
\Pr\left\{\max_j\{s_j^\dag\} - \min_j\{s_j^\dag\} \leq 1\right\} = 0.6003
\end{equation}
where $E\{\cdot\}$ means expectation and $\Pr\{\cdot\}$ means probability of an
event. 
%According to Lemma \ref{lemma:opt_perm_suff} and its remark,
Thus, the permutation in this example is optimal with probability
$0.6003$.

For the best $s$-term approximation ${\bf X}^s$ of a 2D compressible signal
${\bf X}\in \mathbb{R}^{M \times N}$, consider that elements in ${\bf X}^s$
can be divided into several non-overlapped groups, and in each group all
elements share the same probability to be nonzero. Based on the observation at
the beginning of this subsection, a permutation, named {\it group-scan-based
permutation}, can work as follows: 1)~preform group-by-group scan\footnote{That
is, first scan all elements in the first group, then scan all elements in the
second group, ..., and so on.} of the 2D compressible signal ${\bf X}$ into a
vector, and 2)~row-wisely reshape the resulted vector into a new $M\times N$ 2D
signal. In this way, all columns of the best $s$-term approximation of the new
2D signal are likely to have similar sparsity levels.

\begin{mydef}
\label{def:layer_def}
For a 2D signal ${\bf X} \in \mathbb{R}^{M \times N}$, let ${\bf X}(i, j)$ 
denote the element in the position $(i, j)$. The $m$-th ($1\le m< M+N$) layer 
of ${\bf X}$ is the group of all elements ${\bf X}(i, j)$'s satisfying $i+j-1=m$.
\end{mydef}

For example, when $M = N =4$, the following matrix ${\bf X}$
\begin{eqnarray}
{\bf X} =
\begin{bmatrix}
x_1 & x_2 & x_6 & x_7 \\
x_3 & x_5 & x_8 & x_{13} \\
x_4 & x_9 & x_{12} & x_{14} \\
x_{10} & x_{11} & x_{15} & x_{16}
\end{bmatrix}
\label{2D_mtx}
\end{eqnarray}
has 7 layers, including $\{x_1\}$, $\{x_2,x_3\}$, $\{x_4,x_5,x_6\}$, 
$\{x_7,x_8,x_9,x_{10}\}$, $\{x_{11},x_{12},x_{13}\}$, $\{x_{14},x_{15}\}$, 
$\{x_{16}\}$, respectively. The layers of ${\bf X}$ are parallel to each other.

For a 2D compressible signal ${\bf X}$, if elements in each layer of its 
best $s$-term approximation ${\bf X}^s$ have similar {probabilities} to 
be nonzero (an example when this condition is satisfied is to be given 
later in this subsection), then we propose the following zigzag-scan-based 
permutation, which is a special example of the group-scan-based permutation.

Define the \emph{zigzag-scan-based permutation} P:~$\mathbb{R}^{M 
\times N}~\rightarrow~\mathbb{R}^{M \times N}$ for a 2D signal ${\bf X} 
\in \mathbb{R}^{M \times N}$ as P$({\bf X}) = \text{R}(\text{Z}({\bf X}))$, 
where R:~$\mathbb{R}^{MN}~\rightarrow~\mathbb{R}^{M \times N}$ is the 
row-wisely {reshaping} function which row-wisely turns a vector into an 
$M\times N$ matrix and Z:~$\mathbb{R}^{M \times N}~\rightarrow~\mathbb{R}^{MN}$ 
is the zigzag scan function which turns a matrix into a ``zigzag" sequence 
vector.

Correspondingly, define the \emph{inverse zigzag-scan-based permutation} 
P$^{-1}$:~$\mathbb{R}^{M \times N}~\rightarrow~\mathbb{R}^{M \times N}$ for 
a 2D signal ${\bf X}^\dag \in \mathbb{R}^{M \times N}$ as P$^{-1}({\bf X}^\dag) 
= \text{Z}^{-1}(\text{R}^{-1}({\bf X}^\dag))$, where R$^{-1}$:~ $\mathbb{R}^{M 
\times N}~\rightarrow~\mathbb{R}^{MN}$ is a vectorization function which 
row-wisely turns a matrix into a vector and Z$^{-1}$:~$\mathbb{R}^{MN}~
\rightarrow~\mathbb{R}^{M \times N}$ is inverse zigzag scan function which 
turns a ``zigzag" sequence into an $M \times N$ matrix.

For example, the matrix ${\bf X}$ given in (\ref{2D_mtx})
becomes a ``zigzag" sequence after zigzag scan, i.e.,
\[
\text{Z}({\bf X}) = [x_1, x_2, x_3, x_4, x_5, x_6, x_7, \cdots, x_{16}],
\]
and then becomes the permuted signal ${\bf X}^\dag$ after row-wisely reshaping, 
{that is},
\begin{align*}
{\bf X}^\dag
&= \text{P}({\bf X})
= \text{R}(\text{Z}({\bf X}))\\
&= \text{R}([x_1, x_2, x_3, x_4, x_5, x_6, x_7, \cdots, x_{16}])\\
&= \begin{bmatrix}
x_1 & x_2 & x_3 & x_4 \\
x_5 & x_6 & x_7 & x_8 \\
x_9 & x_{10} & x_{11} & x_{12} \\
x_{13} & x_{14} & x_{15} & x_{16}
\end{bmatrix},
\end{align*}
and again becomes a ``zigzag" sequence after vectorization, {that is},
\[
\text{R}^{-1}({\bf X}^\dag) = \text{Z}({\bf X}) = [x_1, x_2, x_3, x_4, x_5,
 x_6, x_7, \cdots, x_{16}],
\]
and then returns to the original signal ${\bf X}$ after inverse zigzag scan, 
i.e., $\text{P}^{-1}({\bf X}^\dag) = \text{Z}^{-1}(\text{R}^{-1} 
({\bf X}^\dag)) = {\bf X}$.

Thus, according to the analysis at the beginning of this subsection, if elements 
in each layer of ${\bf X}^s$ share similar {probabilities} to be nonzero, after 
the zigzag-scan-based permutation, all columns of the permuted ${\bf X}^s$ tend 
to have similar sparsity levels.

\begin{mydef}
\label{def:layer_model}
Consider a 2D compressible signal ${\bf X} \in \mathbb{R}^{M \times N}$ and its 
best $s$-term approximation ${\bf X}^s$.  For given transition layer indices 
$r_0$, $r_1$, $r_2$ and a decay factor $\alpha$, {we say that ${\bf X}$} follows 
the $(r_0, r_1, r_2, \alpha)$-layer model if the probability of the event $E_{m}$ 
that an element in the $m$-th layer of ${\bf X}^s$ is nonzero follows the 
probability distribution
\[
\Pr\left\{E_{m}\right\} = \left\{
\begin{array} {l l }
0 & \quad 1\!\leq m\leq\!r_0 \\
1 & \quad r_0\!+\!1\!\leq m \leq\!r_1 \\
e^{-\!\alpha(m-\!r_0)} & \quad r_1\!+\!1\!\leq m \leq\!r_2 \\
0 & \quad r_2\!+\!1\!\leq m \!\leq\!M\!+\!N\!-\!1.
\end{array} \right.
\]
\end{mydef}

Based on the $(r_0, r_1, r_2, \alpha)$-layer model, we have the following proposition 
for the zigzag-scan-based permutation.

\begin{mypps}
\label{pps:zigzag_perm}
If a 2D compressible signal ${\bf X} \in \mathbb{R}^{M \times N}$ follows the $(r_0, 
r_1, r_2, \alpha)$-layer model with  $r_2 \geq 2r_1-3r_0 - 1$ and $0 \leq r_0 < r_1 
< r_2 \leq \min\{M, N\}$, the zigzag-scan-based permutation $\text{P}(\cdot)$ is an 
acceptable permutation with a large probability that is given as
\begin{align}
&\Pr\left\{\textnormal{P is acceptable}\right\}
= \Pr\left\{||{\bf s}||_\infty > ||{\bf s}^\dag||_\infty\right\} \nonumber &\\
&\geq 1\!-\!\bigg[ \prod_{m=r_1+1}^{r_2} (1-p_m)^m \bigg] \cdot \prod_{j\!=\!1}^{r_2} 
\bigg\{\!1\!+\!\sum_{k\!=\!k_j\!+\!1}^{\substack{\min\{\lceil (r_0+r_2+1) / 2 \rceil, \\
r_2-r_0, r_2-j+1\}}} &\nonumber\\
&\sum_{\substack{a_1\!,a_2\!,\cdots\!,a_{k\!-\!k_j}\!\in\!\mathcal{A}_j \\ a_1\!<\!a_2
\!<\!\cdots\!<\!a_{k\!-\!k_j}}}\!{1 \over (e^{\!\alpha(\!a_1\!-\!1\!-\!r_0\!)}\!-\!1\!)
\!\cdots\!(e^{\!\alpha(\!a_{\!k\!-\!k_j}\!-\!1\!-r_0\!)}\!-\!1)}
\!\bigg\}&
\label{e:propos_1_eq} 
\end{align}

\noindent where ${\bf s}$ and ${\bf s}^\dag$ are the sparsity vectors of the best 
$s$-term approximation of ${\bf X}$ and ${\bf X}^\dag$, respectively, ${\bf X}^\dag$ 
is a 2D signal after the zigzag-scan-based permutation, and for {$1 \leq j \leq r_2$}, 
$\mathcal{A}_j~\overset{\triangle}=~\left\{m_j,~m_j+1,~\cdots~,~r_2\right\}$, $m_j 
= \max\left\{r_1+1, j\right\}$, and
\begin{equation*}
k_j = \left\{
\begin{array} {l l }
r_1-r_0{,} & \quad 1 \leq j \leq r_0 \\
r_1-j+1{,} & \quad r_0+1 \leq j \leq r_1 \\
0{,} & \quad r_1+1 \leq j \leq r_2 .\\
\end{array} \right.
\end{equation*}
\end{mypps}
\begin{IEEEproof}
See Appendix for the proof.
\end{IEEEproof}

\begin{figure}[t]
	\centering
	\includegraphics[width=\textwidth]{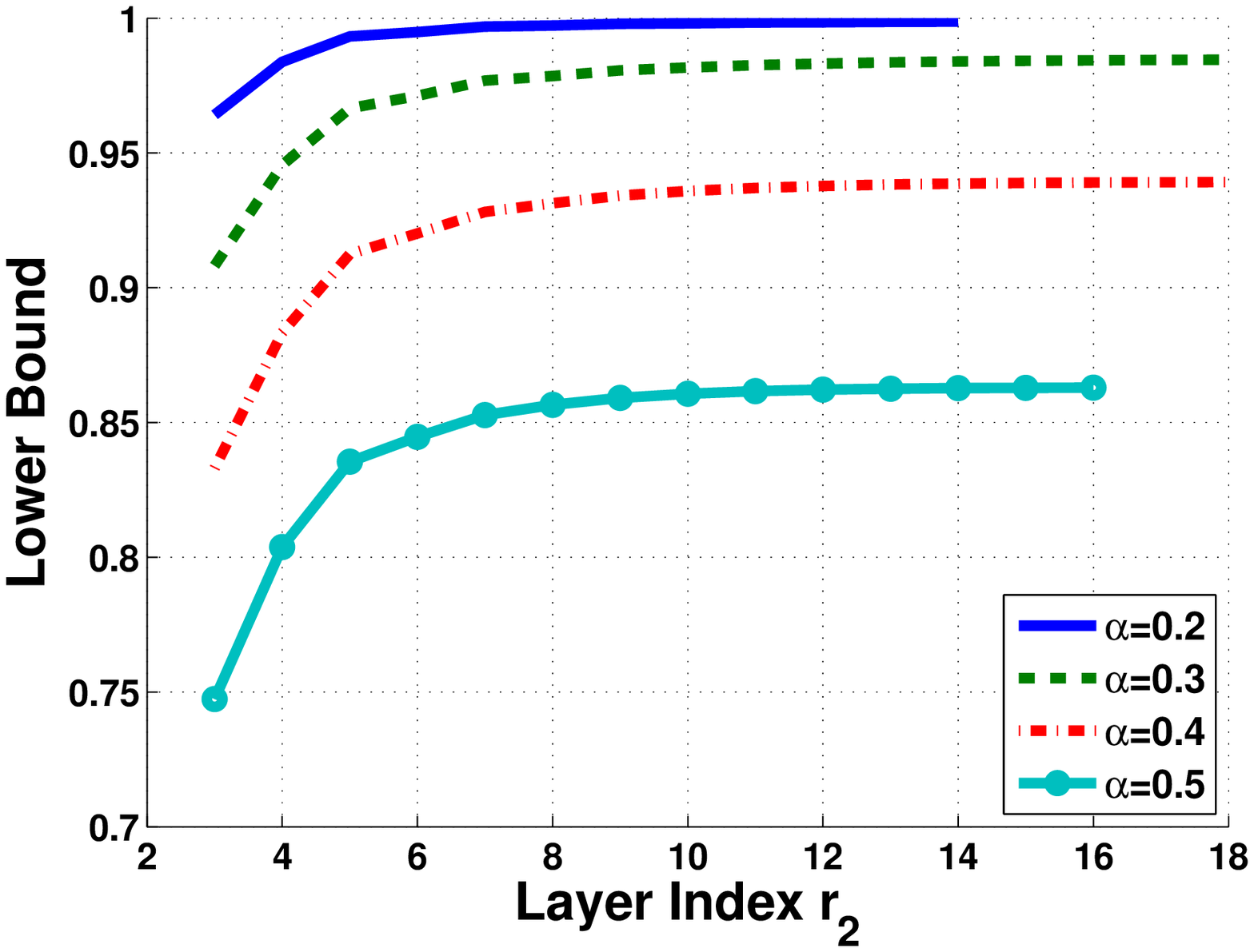}
	\caption{Lower bound of $\Pr\left\{\text{P is acceptable}\right\}$ in 
	(\ref{e:propos_1_eq}) for $r_0~=~0, r_1~=~2$.}
	\label{fig:pps_lb_02}
\end{figure}

\begin{figure}[t]
	\centering
	\includegraphics[width=\textwidth]{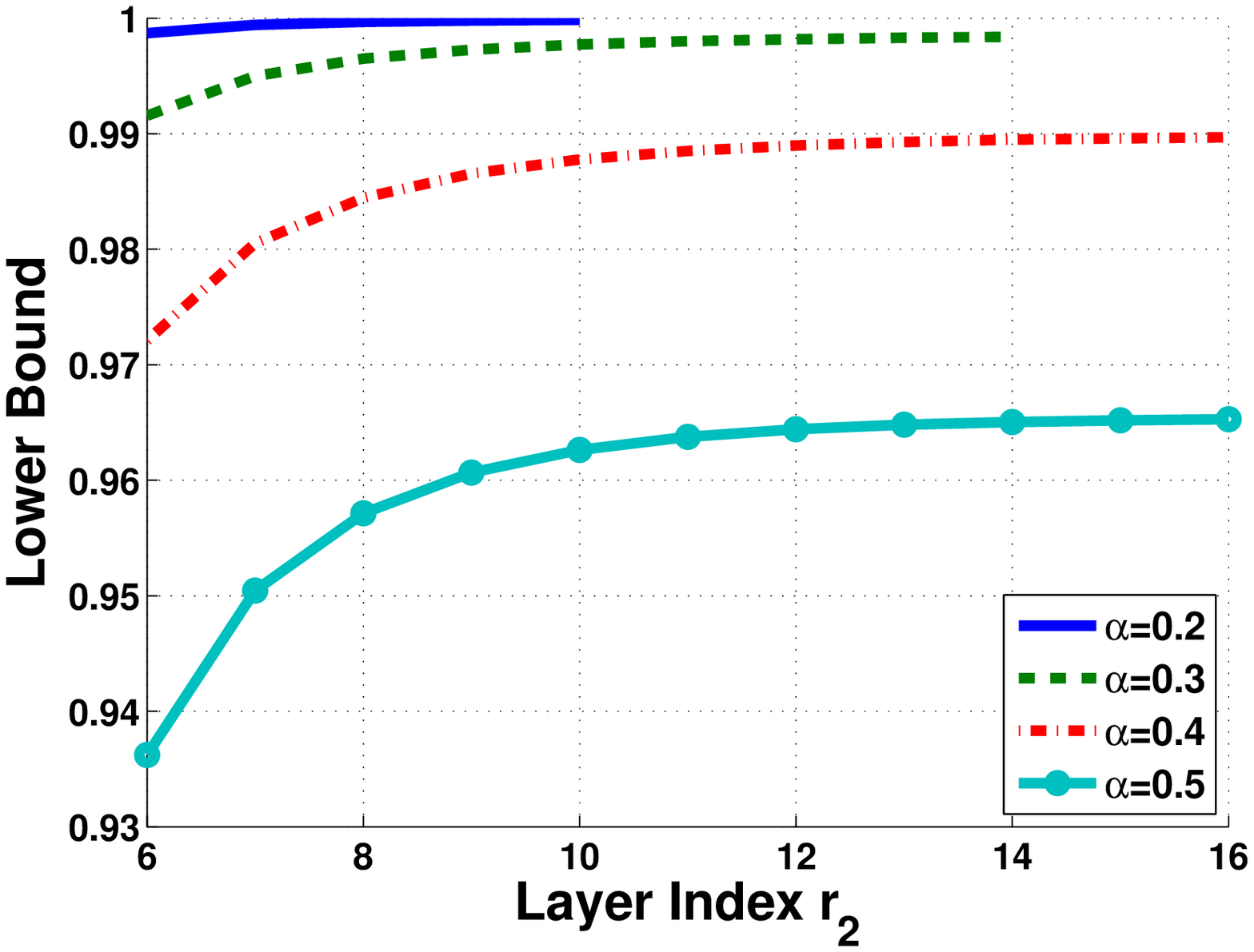}
	\caption{Lower bound of $\Pr\left\{\text{P is acceptable}\right\}$ in 
	(\ref{e:propos_1_eq}) for $r_0~=~3, r_1~=~5$.}
	\label{fig:pps_lb_35}
\end{figure}

Figs.\,\ref{fig:pps_lb_01}--\ref{fig:pps_lb_35} show the value of the lower 
bound on $\Pr$\{P is acceptable\} in (\ref{e:propos_1_eq}) under different 
$\alpha$ and $r_2$ for {1)~$r_0~=~0$, $r_1~=~1$; 2)~$r_0=0, r_1=2$; and 
3)~$r_0=3, r_1=5$.} It can be seen that the lower bound on $\Pr\left\{\text{P 
is acceptable}\right\}$ is large enough in general. For other $r_0$ and $r_1$, 
the {results} are similar.

From Proposition \ref{pps:zigzag_perm}, it can be seen that the zigzag-scan-based 
permutation is an acceptable permutation for a very broad {class of signals. The 
knowledge of exact locations of the nonzero entries {of the best $s$-term 
approximation ${\bf X}^s$, i.e., the knowledge of the support of the 2D signal 
${\bf X}^s$,} is not needed}.

\begin{figure}[t]
	\centering
	\includegraphics[width=\textwidth]{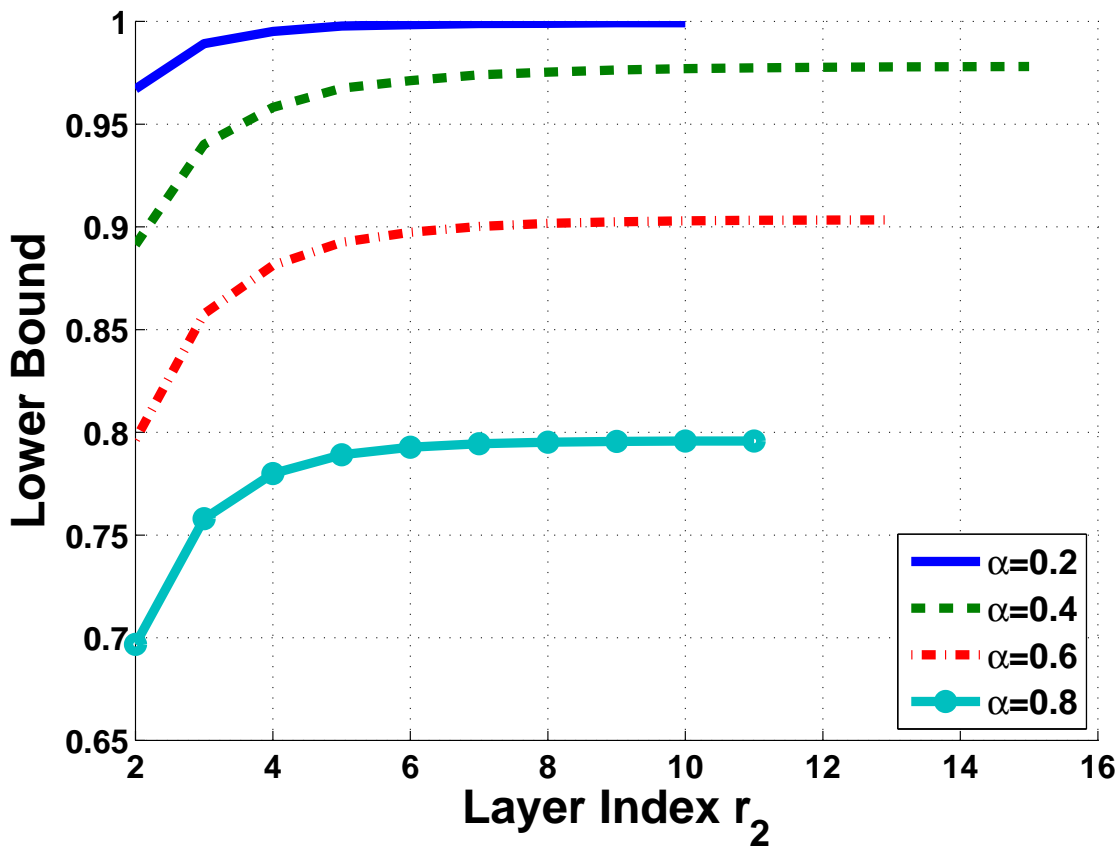}
	\caption{Lower bound of $\Pr\left\{\text{P is acceptable}\right\}$ in 
	(\ref{e:propos_1_eq}) for $r_0~=~0, r_1~=~1$.}
	\label{fig:pps_lb_01}
\end{figure}

\begin{figure}[t]
	\centering
	\begin{subfigure}[b]{0.4\textwidth}
		\centering
		\includegraphics[width=\textwidth]{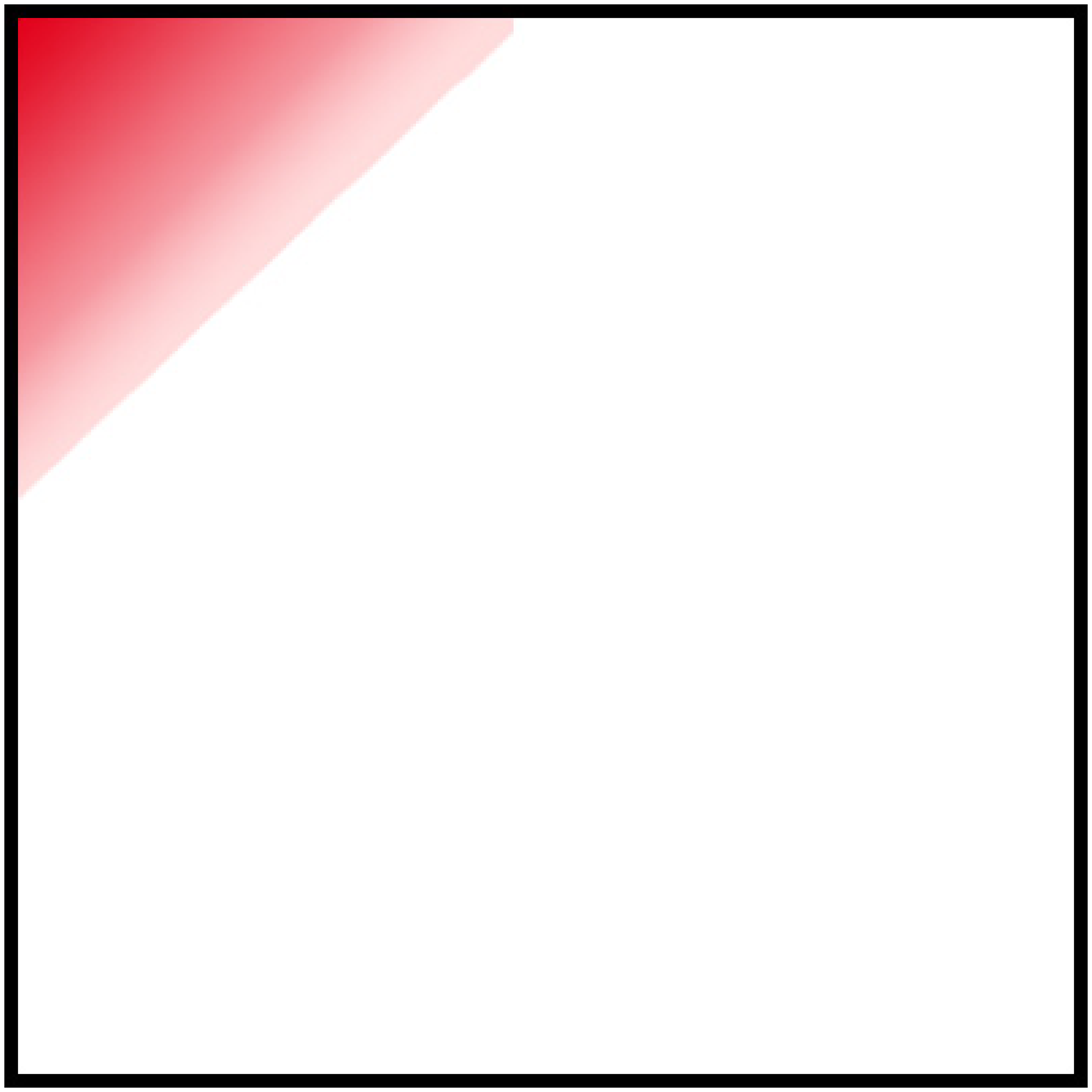}
		\caption{Before the zigzag-scan-based permutation}
		\label{fig:dct2_energy_before_perm}
	\end{subfigure}
	\quad
	\begin{subfigure}[b]{0.4\textwidth}
		\centering
		\includegraphics[width=\textwidth]{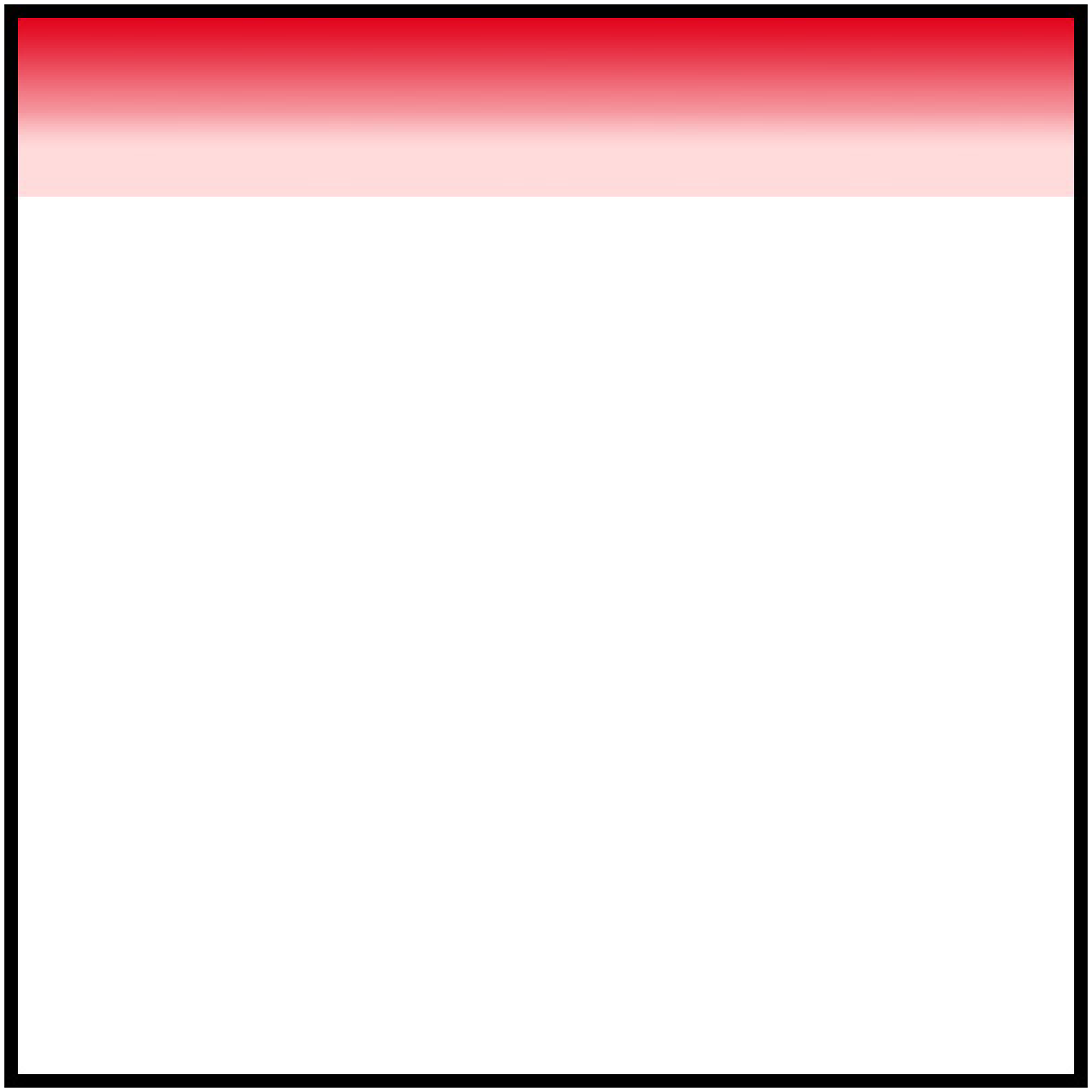}
		\caption{After the zigzag-scan-based permutation}
		\label{fig:dct2_energy_after_perm}
	\end{subfigure}
	\caption{Energy distribution of a DCT2 coefficient matrix before and 
	after the zigzag-scan-based permutation.}
	\label{fig:dct2_energy}
\end{figure}

As an example, we show that the zigzag-scan-based permutation is particularly 
useful for 2D discrete cosine transform (DCT2) coefficient matrices of 2D 
piecewise smooth image signals. {Since the DCT2 coefficient matrix of a 
piecewise smooth image signal typically has most of its large elements lie 
in the top left corner, and small elements lie in the bottom right corner 
because most of its energy is concentrated in low frequencies,} the zigzag 
scan process is commonly used in image compression like JPEG \cite{Wallace1992}. 
Thus, the DCT2 coefficient matrices of piecewise smooth image signals satisfy 
the $(r_0, r_1, r_2, \alpha)$-layer model with $r_0=0$ (which will also be 
shown via simulation in Subsection \ref{sim_dct2_zigzag}), and thus, the 
proposed zigzag-scan-based permutation has a large probability to be an 
acceptable permutation when parallel CS is applied to the DCT2 coefficient 
matrices. Note that {the knowledge of the layer indices $r_1$, $r_2$ and the 
decay factor $\alpha$ of the layer model is not needed} when applying the 
zigzag-scan-based permutation {to} the DCT2 coefficient matrices.

\figurename\,\ref{fig:dct2_energy} shows the difference before and after the 
zigzag-scan-based permutation when the 2D signal is the DCT2 coefficient matrix 
of an image. The energy, which can be loosely viewed as an interpretation of 
the sparsity vector, if all non-zero entries of the 2D signal have magnitude 
of the same order, is distributed more evenly among columns after the 
zigzag-scan-based permutation.

One advantage of the zigzag-scan-based permutation is that it is a pre-defined 
permutation, and thus, the encoder and decoder know it in advance without any 
additional communication. In Subsection \ref{sim_dct2_zigzag}, we {will} also 
show by simulation that the zigzag-scan-based permutation is an acceptable
permutation for DCT2 coefficient matrices of several typical images.

\section{Example of Video Compression via Parallel CS with Permutations in 
Wireless Multimedia Sensor Networks}
\label{cvs}
As an application example, we design a pair of CS video encoder and
decoder based on parallel CS with the zigzag-scan-based permutation.
This CS video encoder and decoder can be plugged into the application layer of
the compressive distortion minimizing rate control (C-DMRC) system
\cite{Pudlewski2012}.
In wireless multimedia sensor networks, the C-DMRC system is preferred compared
to traditional video coding standards such as MPEG, H.264, since the C-DMRC
system has less-complex video encoder and can tolerate much higher bit error
rates.
The CS video encoder and decoder in the C-DMRC system are built based on the
block CS architecture proposed in \cite{Gan2008}. 
Thus, as we discussed in Section~\ref{sec:intro}, the CS video decoder in the
C-DMRC system requires a joint reconstruction.
By replacing the CS video encoder and decoder at the application layer of the
C-DMRC system with the CS video encoder and decoder based on parallel CS
architecture, the computational complexity of the video decoder can be reduced
and the reconstruction process can be parallelized. 

\begin{figure*}
  \centering
  \begin{subfigure}[b]{0.8\textwidth}
    \centering
	\includegraphics[width=\textwidth]{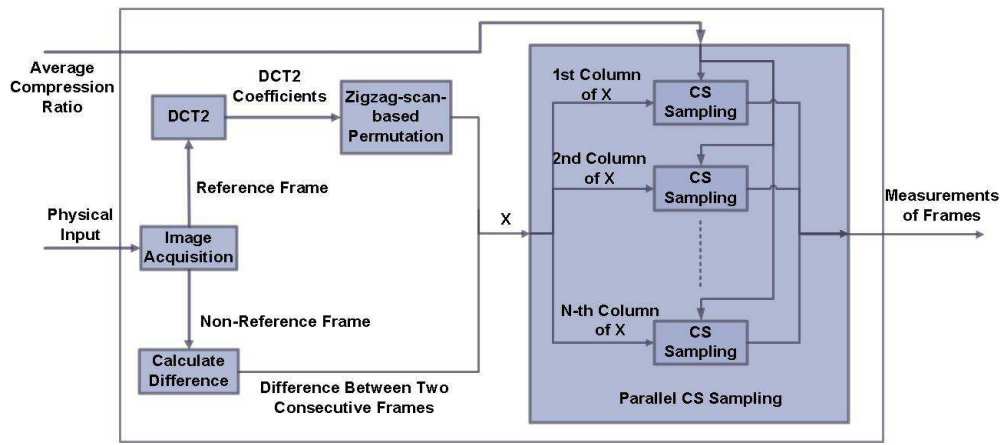}
	\caption{Block diagram of the CS video encoder.}
	\label{fig:csv_encoder_block_diag}
      \end{subfigure}
      \\
      \begin{subfigure}[b]{0.8\textwidth}
	\centering
	\includegraphics[width=\textwidth]{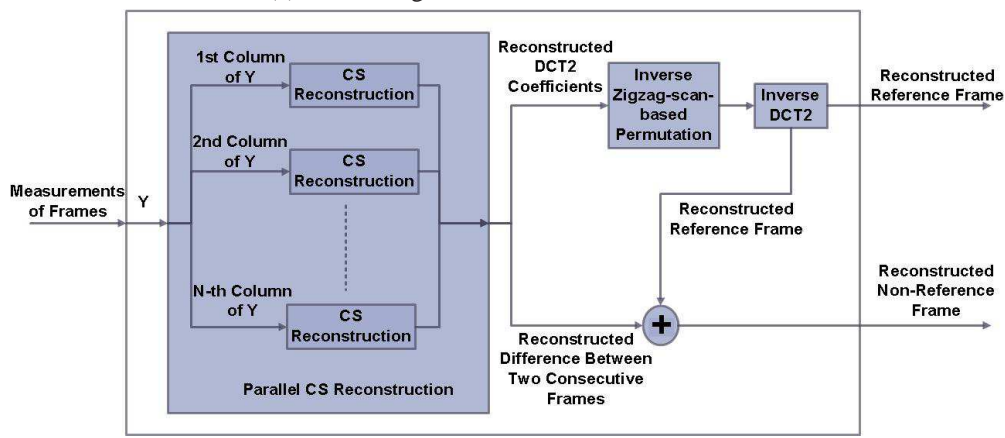}
	\caption{Block diagram of the CS video decoder.}
	\label{fig:csv_decoder_block_diag}
      \end{subfigure}
      \caption{Block diagram of the CS video encoder and decoder.}
\end{figure*}
In this example, frames with odd indices and even indices are taken as 
reference frames and non-reference frames, respectively\footnote{More 
sophisticated index assignment schemes for the reference frame and 
non-reference frame can be used as well.}. The block diagram of the CS 
video encoder is shown in \figurename\,\ref{fig:csv_encoder_block_diag}.
The average compression ratio is computed by the rate controller at the
transport layer according to current network status (e.g., the
end-to-end round trip time and the estimated sample loss rate of the network),
and it controls the number of measurements for a video frame.
For every pair of a reference frame and its following non-reference frame, the
rate controller gives an average compression ratio. 
According to this average compression ratio, the compression ratios of the
reference and non-reference frames in a pair are obtained. 
At the output of the CS video encoder we have the frame measurements.
The image acquisition device turns the physical input into video frames and
outputs the video frames to different processing blocks according to the frame
index.

The procedure for encoding the reference frame is as
follows:~1)~compute DCT2 on the reference frame;
2)~perform the zigzag-scan-based permutation on the DCT2 coefficient matrix; 
3)~perform parallel compressed sampling of the permuted DCT2 coefficient matrix. 
The procedure for encoding the non-reference frame is as follows:~1)~compute the
difference between the non-reference frame and the preceding reference frame;
2)~perform parallel compressed sampling of this difference.
The outputs of all CS sampling processors are
combined.\footnote{Quantization is omitted in the
example presented here, but it has to been done in a practical video coding
scenario.} 
For the non-reference frames, no permutation is performed since the difference
between two consecutive frames, especially in videos with slow motion,
is sparse enough so that the sparsity level of each column is too
small to have significant difference from column to column. 
Thus, the permutation does not bring significant improvement, which
we have checked by simulations in \cite{Fang2012Asilomar}.

Considering that the sparsity level of the difference between a 
non-reference frame and its preceding reference frame is smaller than that of
the DCT2 coefficient matrix of the reference frame, the compression ratio of
the non-reference frames should be higher than that of the reference frames,
i.e., fewer measurements are assigned to the non-reference frames. 
In our experiment in Section~\ref{sim}, we set the ratio of measurements being
4:1, i.e., the number of measurements for reference frames is 4 times that
for non-reference frames. 
For example, if current average compression ratio given by the rate controller
is 0.5, then the compression ratio of the reference frame is 0.8 and the
compression ratio of the non-reference frame is 0.2. 
Other ratios can be set according to the motion intensity of the video.

The block diagram of the CS video decoder is shown in
\figurename\,\ref{fig:csv_decoder_block_diag}. 
To decode a reference frame at the receiver side, the following steps are
performed:~1)~perform parallel CS reconstruction of the permuted DCT2 coefficient 
matrix from the measurements of the reference frame; 2)~perform the inverse 
zigzag-scan-based permutation on the reconstructed permuted DCT2 coefficient
matrix; 3)~perform inverse DCT2 on the reconstructed DCT2 coefficient 
matrix. To decode a non-reference frame, the following steps are performed:
1)~perform parallel CS reconstruction of the difference between the non-reference 
frame and its preceding reference frame from the measurements of the non-reference 
frame;  2)~add the reconstructed difference between the non-reference frame 
and its preceding reference frame to the corresponding reconstructed reference
frame. For parallel CS reconstruction, any $\ell_1$-norm minimization solver, 
e.g., the basis pursuit algorithm, can be used.

\section{Simulation Results}
\label{sim}
\subsection{The Layer Model and the Zigzag-scan-based Permutation}
\label{sim_dct2_zigzag}

\begin{figure}
	\centering
	\includegraphics[width=\textwidth]{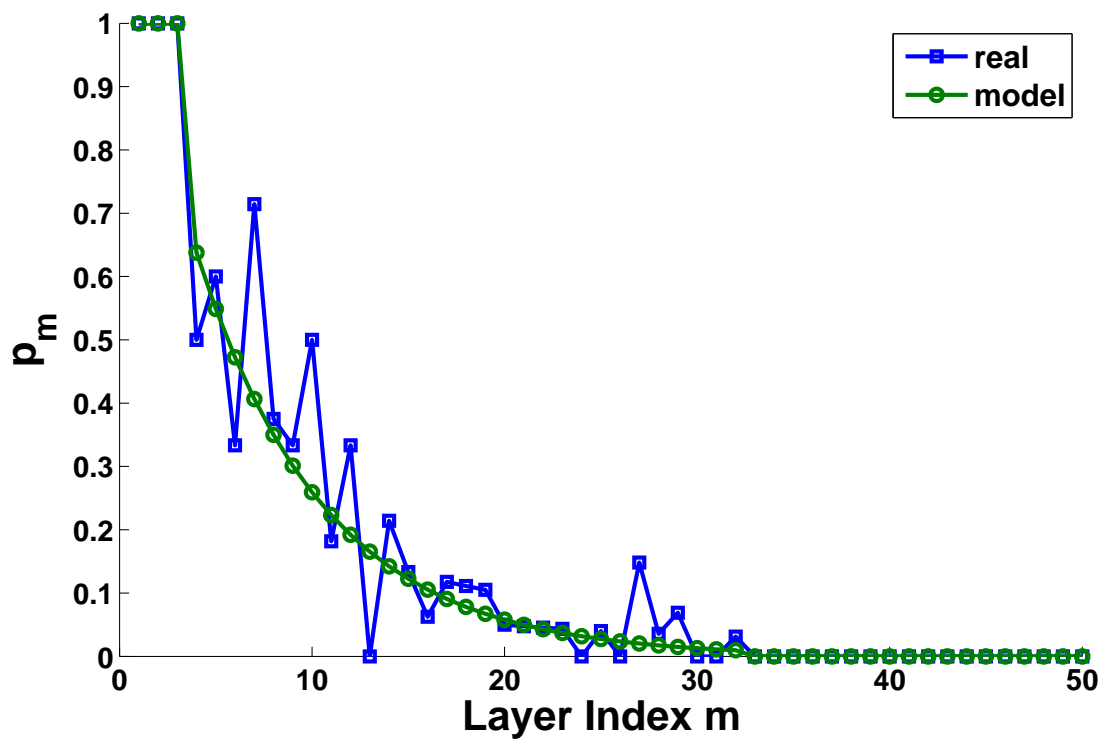}
	\caption{Layer model of Boat.tiff.}
	\label{fig:layer_model_boat}
\end{figure}

We first check the layer model for the DCT2 coefficient matrix of the gray
image: Boat (512 $\times$ 512). The format used in the simulation is tagged
image file format (TIFF). The best $s$-term approximation is obtained by keeping
all DCT2 coefficients with {magnitudes} not less than 1000 and changing the
remaining to zeros. In \figurename\,\ref{fig:layer_model_boat}, the x-axis is
the layer index $m$, and y-axis is the probability of an element in the $m$-th
layer of the best $s$-term approximation ${\bf X}^s$ of the DCT2 coefficient
matrix to be nonzero, calculated as {$p_m = (1/m)\sum_{i+j-1=m} \text{I}({\bf
X}^s(i, j) \neq 0)$, where} $\text{I}(\cdot)$ is the indicator function. The
{$p_m$'s versus layer index $m$ for} the real image ``Boat.tiff" and the result
of the $(r_0, r_1, r_2, \alpha)$-layer model with $r_0~=~0, r_1~=~3, r_2~=~32,
\alpha = 0.15$ are shown in \figurename\,\ref{fig:layer_model_boat}. It can be
seen that the two curves are close to each other. Similar results are also
achieved for other images. Then according to Proposition \ref{pps:zigzag_perm},
the zigzag-scan-based permutation {is} an acceptable permutation {for DCT2
coefficient matrices of such images} with an overwhelming probability.

\begin{table}
\caption{Comparison of $||{\bf s}||_\infty$ before and after the
zigzag-scan-based permutation.}
\label{table:zigzag_perm}
\centering
\begin{tabular}{c||c|c|c|c}
\hline\hline
\multirow{2}{*}{{\bf Image}}
 & \multicolumn{4}{c}{\rule[-1ex]{0pt}{3.5ex}{\bf Magnitude Threshold}} \\
 \cline{2-5}\rule[-1ex]{0pt}{3.5ex}
 & 400 & 600 & 800 & 1000 \\ %[0.5ex]
\hline\rule[-1ex]{0pt}{3.5ex}
Boat & 33 vs. 2 & 23 vs. 2 & 19 vs. 2 & 16 vs. 1 \\ %[0.5ex]
\hline\rule[-1ex]{0pt}{3.5ex}
Cameraman & 13 vs. 2 & 8 vs. 2 & 7 vs. 1 & 4 vs. 1 \\ %[0.5ex]
\hline\rule[-1ex]{0pt}{3.5ex}
Lena & 14 vs. 3 & 11 vs. 2 & 8 vs. 1 & 7 vs. 1 \\ %[0.5ex]
\hline\rule[-1ex]{0pt}{3.5ex}
Peppers & 27 vs. 3 & 15 vs. 2 & 11 vs. 2 & 11 vs. 2 \\ %[0.5ex]
\hline\hline
\end{tabular}
\end{table}
{The changes of $||{\bf s}||_\infty$} of the best $s$-term approximation 
of the DCT2 coefficient matrix before and after the zigzag-scan-based 
permutation are shown in Table\,\ref{table:zigzag_perm}. The DCT2 coefficient 
matrices are {taken} from four test images: Boat (512 $\times$ 512), Lena 
(512 $\times$ 512), Cameraman (256 $\times$ 256), Peppers (512 $\times$ 512). 
The best $s$-term approximation is chosen according to different magnitude 
thresholds, i.e., keeping DCT2 coefficients whose {magnitudes} are not less 
than the magnitude threshold and setting the remaining to be zeros. 
Table\,\ref{table:zigzag_perm} shows that $||{\bf s}||_\infty$ decreases 
significantly after {the} zigzag-scan-based permutation, which is 
consistent with Proposition \ref{pps:zigzag_perm}.

\subsection{Image Compression via Parallel CS with the Zigzag-Scan-Based 
Permutation}
\label{sim_img_pcs}
\begin{figure}[t]
	\centering
	\includegraphics[width=\textwidth]{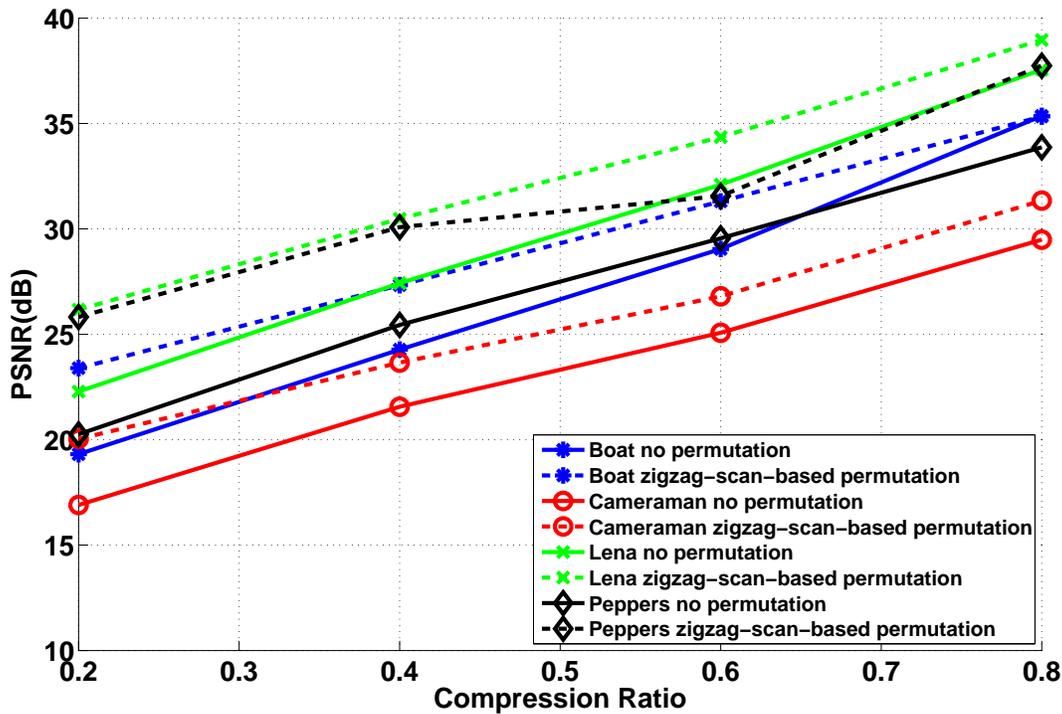}
	\caption{PSNR for the parallel CS scheme with/without the zigzag-scan-based 
	permutation in image compression.}
	\label{fig:psnr}
\end{figure}

The performance of image compression via parallel CS with the zigzag-scan-based
permutation is shown by compressing the DCT2 coefficients of four images: Boat,
Lena, Cameraman and Peppers. The PSNR is employed to show the reconstruction
performance. We compare the performances of the parallel CS scheme for {the
configurations}:~1)~with no permutation{;} 2)~with the zigzag-scan-based
permutation. Entries of the sensing matrix ${\bf A} \in \mathbb{R}^{K \times M}$
are drawn from Gaussian ensembles, with variance being $1/K$. The parallel CS
reconstruction is implemented using basis pursuit algorithm by the CVX
optimization toolbox.\footnote{The toolbox was downloaded at
http://cvxr.com/cvx.} Other reconstruction algorithms {than} the basis pursuit
can also be used. PSNR performance for different methods is shown in
\figurename\,\ref{fig:psnr} {versus} the compression ratio, which is the ratio
of the number of measurements to the total number of elements in the DCT2
coefficient matrix.

From \figurename\,\ref{fig:psnr}, we can see that under the same compression 
ratio, the zigzag-scan-based permutation helps to improve the PSNR by around 
{4~dB} for all images. Consequently, it shows that the PNSR performance is 
indeed improved significantly after permutation.

\subsection{Video Compression via Parallel CS with the Zigzag-scan-based 
Permutation}
\begin{figure}
	\centering
    \includegraphics[width=\textwidth]{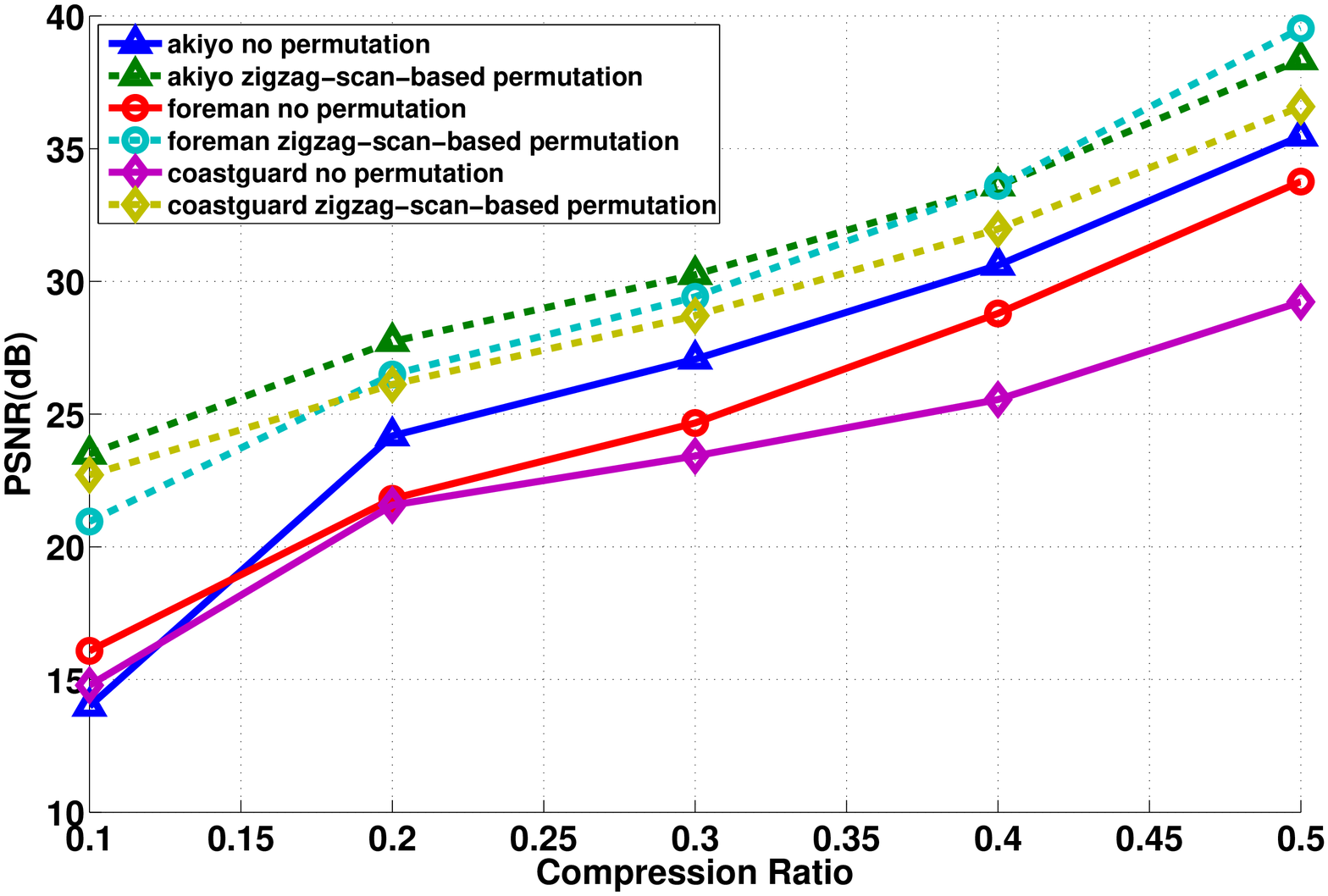}
	\caption{Average PSNR of reconstructed reference frames.}
	\label{fig:ref}
\end{figure}

\begin{figure}
	\centering
    \includegraphics[width=\textwidth]{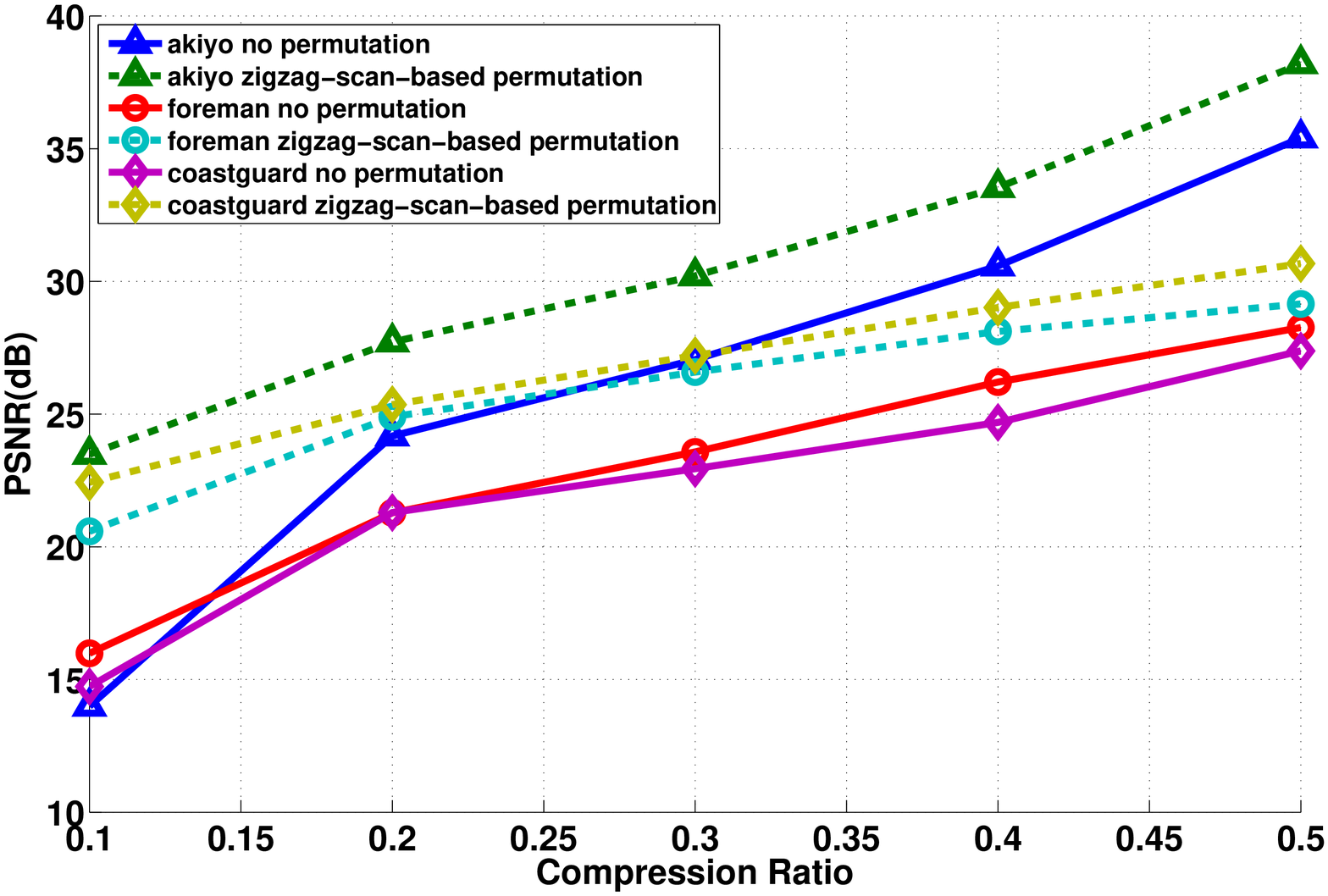}
	\caption{Average PSNR of reconstructed non-reference frames.}
	\label{fig:nonref}
\end{figure}

The test video sequences in this example are three standard YUV video sequences:
Akiyo, Foreman, Coastguard. 
The format used in the simulation is quarter common intermediate format (QCIF).
The performance of the proposed video compression scheme is shown by compressing
the luminance components of the first 10 frames, i.e., 5 reference frames and 5
non-reference frames.
The average PSNR for reference frames and non-reference frames is used as
performance metric.
All settings are the same as in the example in Subsection~\ref{sim_img_pcs}.
PSNR performance for different methods is shown in
Figs.\,\ref{fig:ref}~and~\ref{fig:nonref} versus the average compression ratio,
that is, $(\text{compression ratio of reference frames} + \text{compression
ratio of non-reference frames})/2$.

From \figurename\,\ref{fig:ref}, we can see that under the same average
compression ratio, the zigzag-scan-based permutation helps to improve the PSNR
of reference frames by around 3$\sim$9~dB for Akiyo, 5$\sim$6~dB for Foreman and
4$\sim$8~dB for Coastguard. 
\figurename\,\ref{fig:nonref} shows that the zigzag-scan-based permutation also
improves the PSNR performance of non-reference frames by around 3$\sim$9~dB for
Akiyo, 2$\sim$5~dB for Foreman and 3$\sim$7~dB for Coastguard.
The improved PSNR for the non-reference frame is a bit lower than
that of the preceding reference frame because the reconstruction of the non-reference frame
relies on both the reconstruction of its preceding reference frame and the reconstruction of
difference between the non-reference frame and its preceding reference frame.

To show the advantage of the video compression scheme proposed in
Section~\ref{cvs}, we compare the total time of reconstructing one pair of
reference and non-reference frames using (i)~the video encoder and decoder 
employed in the C-DMRC system proposed in \cite{Pudlewski2012} and (ii)~the video 
encoder and decoder proposed in Section~\ref{cvs}. We also show the PSNRs of the 
reconstructed reference and non-reference frames for both schemes. The video 
sequence used in the simulation is the standard YUV sequence Akiyo (QCIF format). 
The measurement matrices used in the C-DMRC system and our scheme are the 
scrambled block Hadamard matrix (block length equals to 32) and the random 
Gaussian matrix, respectively. The CS reconstruction algorithm is implemented 
using the $l_1$-magic package.\footnote{The package is available at
http://users.ece.gatech.edu/~justin/l1magic.} To eliminate the effects of 
randomness, we run 200 trials for each average compression ratio and show the 
average PSNR and total reconstruction time. The results are shown in
Tables~\ref{table:video_sbhe}~and~\ref{table:video_pcs}.

\begin{table}
\caption{Total reconstruction time and PSNR of reconstructed video frames using
  the video encoder and decoder employed in \cite{Pudlewski2012}.}
\label{table:video_sbhe}
\centering
\begin{tabular}{c|c||c|c|c|c|c} 
  \hline\hline
  \multicolumn{2}{c||}{\rule[1.5ex]{0pt}{3.5ex}\parbox{5cm}{\centering{\bf Average
  Compression Ratio}}}
  & 0.1 & 0.2 & 0.3 & 0.4 & 0.5 \\ [3ex]
  \hline
  \multicolumn{2}{c||}{\rule[0.5ex]{0pt}{3.5ex}\parbox{5cm}{\centering{\bf Reconstruction Time
  (seconds)}}} & 55.32 & 47.34 & 37.23 & 37.08 & 30.49 \\ [2ex]
  \hline\rule[1ex]{0pt}{3.5ex}
  \multirow{2}{0.8cm}[-3ex]{\centering{\bf PSNR (dB)}} 
  & \parbox{2.2cm}{\centering{\bf Reference Frame}} 
  & 24.43 & 27.52 & 29.79 & 32.53 & 36.24 \\ [2ex]
  \cline{2-7}\rule[1.5ex]{0pt}{3.5ex}
  & \parbox{2.2cm}{\centering{\bf Non-reference Frame}} 
  & 24.44 & 27.53 & 29.73 & 32.27 & 35.32\\ [3ex]
  \hline\hline
\end{tabular}
\end{table}

\begin{table}
\caption{Total reconstruction time and PSNR of reconstructed video frames using
  the video encoder and decoder proposed in Section~\ref{cvs}.}
\label{table:video_pcs}
\centering
\begin{tabular}{c|c||c|c|c|c|c} 
  \hline\hline
  \multicolumn{2}{c||}{\rule[1.5ex]{0pt}{3.5ex}\parbox{5cm}{\centering{\bf Average
  Compression Ratio}}}
  & 0.1 & 0.2 & 0.3 & 0.4 & 0.5 \\ [3ex]
  \hline
  \multicolumn{2}{c||}{\rule[0.5ex]{0pt}{3.5ex}\parbox{5cm}{\centering{\bf
  Reconstruction Time (seconds)}}}
  & 12.85 & 14.30 & 20.17 & 18.40 & 18.67 \\ [2ex]
  \hline\rule[1ex]{0pt}{3.5ex}
  \multirow{2}{0.8cm}[-3ex]{\centering{\bf PSNR (dB)}} 
  & \parbox{2.2cm}{\centering{\bf Reference Frame}}
  & 24.17 & 27.30 & 30.32 & 33.79 & 38.34 \\ [2ex]
  \cline{2-7}\rule[1.5ex]{0pt}{3.5ex}
  & \parbox{2.2cm}{\centering{\bf Non-reference Frame}}
  & 24.17 & 27.29 & 30.29 & 33.71 & 38.10 \\ [3ex]
  \hline\hline
\end{tabular}
\end{table}

It can be seen from Tables~\ref{table:video_sbhe}~and~\ref{table:video_pcs} that 
the reconstruction time using the video encoder and decoder proposed in 
Section~\ref{cvs} is less than that for the video encoder and decoder employed in
\cite{Pudlewski2012}, especially when the compression ratio is low. In addition, 
if there are multiple decoding processors simultaneously reconstructing the columns of a video 
frame as shown in \figurename\,\ref{fig:csv_decoder_block_diag}, the reconstruction 
time can be further reduced approximately to the total reconstruction time divided 
by the number of decoding processors. It can also be observed in Table~\ref{table:video_sbhe} 
that the time for reconstruction using the video encoder and decoder employed in
\cite{Pudlewski2012} decreases as the average compression ratio increases. This is 
because the reconstruction algorithm converges faster as the number of measurements 
increases. According to Table~\ref{table:video_pcs}, the time for reconstruction 
using the video encoder and decoder proposed in Section~\ref{cvs} is less sensitive 
to the compression ratio. In addition, we can see that compared to the video encoder 
and decoder employed in \cite{Pudlewski2012}, the PSNR of reconstructed video frames 
for the video encoder and decoder proposed in Section~\ref{cvs} is larger when the
average compression ratio is larger than 0.3, and is almost the same (less than
0.3dB degradation) when the average compression ratio is smaller than 0.3.

\section{Conclusion and Discussion}
\label{conclusion}
A parallel CS scheme with permutation has been proposed. It has been proved that 
with a so-called acceptable permutation, the RIP condition for the sensing matrix 
in the parallel CS can be relaxed. {The group-scan-based permutation has been 
introduced.} As an example, the zigzag-scan-based permutation for 2D signals 
satisfying the $(r_0, r_1, r_2, \alpha)$-layer model, such as DCT2 coefficient 
matrices of 2D images, has been analyzed. The application {to} image and video 
compression has been discussed as well. In the simulations, it has been shown 
that the zigzag-scan-based permutation for DCT2 coefficient matrices of images 
is an acceptable permutation. In addition, the simulation results have shown 
that the proposed scheme improves the reconstruction performance of images and 
videos in terms of PSNR significantly.

Finally, it is worth mentioning that the zigzag-scan-based permutation is 
{designed} for signals satisfying the {proposed} layer model. {If} a signal 
has most of its large entries clustered around one or more fixed locations, 
the more general group-scan-based permutation {is applicable}. {Similarly to 
the zigzag-scan-based permutation for the layer model, a lower bound on the 
probability that the group-scan-based permutation} is an acceptable permutation 
{can} be derived given a mathematical model for the distribution pattern of 
large entries in the signal.

\section{Appendix: Proof of Proposition \ref{pps:zigzag_perm}}
\begin{IEEEproof}
Denote the $j$-th element of the sparsity vector ${\bf s}$ as $s_j$, i.e., 
the sparsity level of the $j$-th column of ${\bf X}^s$ is $s_j$. Since 
${\bf X}$ follows the $(r_0, r_1, r_2, \alpha)$-layer model, the nonzero 
elements in ${\bf X}^s$ are all in layers of ${\bf X}^s$ whose indices range 
from $r_0+1$ to $r_2$. After performing the zigzag-scan-based permutation on 
${\bf X}^s$, the maximal number of nonzero entries in any column is $u {=} 
\lceil (r_0+r_2+1)(r_2-r_0)/2N \rceil$. Therefore, $u \geq ||{\bf s}^\dag
||_\infty$. Let $l = \lceil (r_0+r_2+1) / 2 \rceil$. Since $r_2 \leq 
\min~\{M,~N\}$ and $r_2 \geq 2r_1 - 3r_0 - 1$, we have $l \geq u$ and $l 
\geq r_1-r_0$.

As a result, the probability that the zigzag-scan-based permutation 
of a 2D signal satisfying the $(r_0, r_1, r_2, \alpha)$-layer model 
is an acceptable permutation can be expressed as
\begin{subequations}
\begin{align}
&\Pr\left\{\text{P is acceptable}\right\}
\label{eq:app_1} = \Pr\{||{\bf s}||_\infty > ||{\bf s}^\dag||_\infty\} \\
\label{eq:app_2} &= \sum_{t=1}^{r_2-r_0} \Pr\{||{\bf s}||_\infty = t, 
||{\bf s}^\dag||_\infty \leq t-1\} \\
\nonumber &\geq \sum_{t=u+1}^{r_2-r_0} \Pr\{||{\bf s}||_\infty = t, 
||{\bf s}^\dag||_\infty \leq t-1\} \\
\label{eq:app_3} &= \sum_{t=u+1}^{r_2-r_0} \Pr\{||{\bf s}||_\infty 
= t\} \\
\nonumber &= \Pr\{||{\bf s}||_\infty \geq u+1\} {= 1 - \Pr\{||
{\bf s}||_\infty \leq u\}}\\
&\geq 1 - \Pr\{||{\bf s}||_\infty \leq l\} \label{eq:app_3_1}.
\end{align}
\end{subequations}
{To derive} (\ref{eq:app_1}), we have used the fact that an acceptable 
permutation must result in $||{\bf s}^\dag||_\infty < ||{\bf s}||_\infty$. 
For deriving (\ref{eq:app_2}), we have used the fact that the maximal 
sparsity level among columns of the best $s$-term approximation ${\bf X}^s$ 
is upper bounded by $(r_2-r_0)$, i.e., $||{\bf s}||_\infty \leq r_2 - r_0$, 
which immediately follows from the $(r_0, r_1, r_2, \alpha)$-layer model. 
For deriving (\ref{eq:app_3}), we have used the fact that $u \geq 
||{\bf s}^\dag||_\infty$. Finally, for deriving (\ref{eq:app_3_1}), we 
have used the fact that $u \leq l$. Based on (\ref{eq:app_3_1}), we 
{focus on} the cumulative distribution function of $||\mathbf{s}||_\infty$.

\begin{figure}
	\centering
	\includegraphics[width=0.5\textwidth]{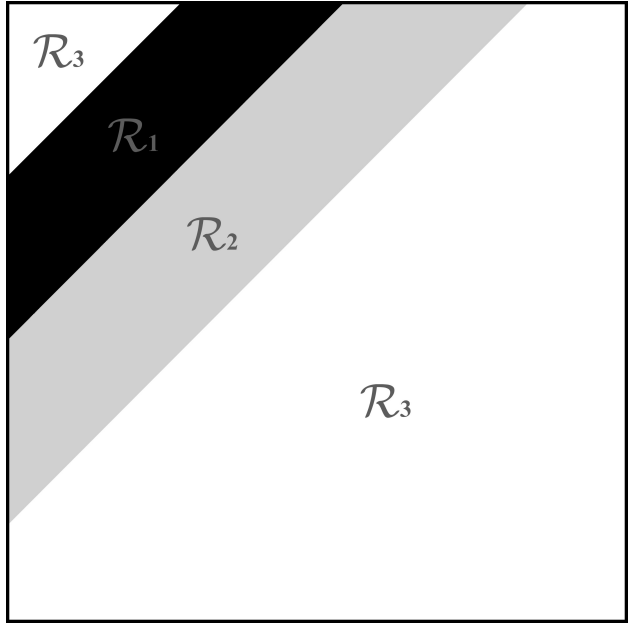}
	\caption{Regions in the $(r_0, r_1, r_2, \alpha)$-layer model.}
	\label{fig:region}
\end{figure}

Since the events {that $s_j \leq l$} for different $j$'s are independent 
with each other, we have
\begin{equation}
\label{eq:app_4}
\Pr\{||{\bf s}||_\infty \leq l\}
= \prod_{j=1}^N \Pr\{s_j \leq l\}.
\end{equation}
{Moreover, since} the position $(i, j)$ of an element in ${\bf X}^s$ 
indicates the index $m$ of the layer where the element is located, i.e., 
$m~=~i+j-1$, we can define three regions in ${\bf X}^s$:
\begin{eqnarray*}
\mathcal{R}_1 &=& \{(i, j) \in \mathbb{Z}^2| r_0+1 \leq i+j-1 \leq r_1 \} \\
\mathcal{R}_2 &=& \{(i, j) \in \mathbb{Z}^2| r_1+1 \leq i+j-1 \leq r_2 \} \\
\mathcal{R}_3 &=& \{(i, j) \in \mathbb{Z}^2| 1 \leq i+j-1 \leq r_0 \} \cup \\
&& \{(i, j) \in \mathbb{Z}^2| r_2+1 \leq i+j-1 \leq M+N-1\}{.}
\end{eqnarray*}
{These regions are separated by three transition layers,} i.e., the 
$r_0$-th layer, the $r_1$-th layer and the $r_2$-th layer{, and they 
are} shown in \figurename\,\ref{fig:region}. Therefore, according to 
Definition \ref{def:layer_model}, all elements of ${\bf X}^s$ are 
nonzero with probability 1 in region $\mathcal{R}_1$. In region 
$\mathcal{R}_3$, all elements of ${\bf X}^s$ are zero with probability 
1. In region $\mathcal{R}_2$, the probability of an element to be 
nonzero decreases exponentially with decay factor $\alpha$ as the 
layer index $m$ increases. We use $p_m$ to denote the probability of 
{an element in the $m$-th layer of ${\bf X}^s$} to be nonzero.

For the $j$-th column of ${\bf X}^s$, if $r_2+1 \leq j \leq N$, all 
elements of the column are in region $\mathcal{R}_3$ and thus are 
all zeros. We have $\Pr\{s_j \leq l\} = 1$ since $l \geq r_1-r_0 
\geq 1$. According to (\ref{eq:app_4}), we have
\begin{equation}
\label{eq:app_5}
\Pr\{||{\bf s}||_\infty \leq l\}
= \prod_{j=1}^{r_2} \Pr\{s_j \leq l\}
= \prod_{j=1}^{r_2} \sum_{k=0}^l \Pr\{s_j\!=\!k\}.
\end{equation}
{Consequently}, we focus on the probability distribution of $s_j$ 
for the first $r_2$ columns of ${\bf X}^s$, i.e., $\Pr\{s_j = k\}$ 
for all $0 \leq k \leq l$ and $1 \leq j \leq r_2$.

Let $k_j$ denote the number of elements in the $j$-th column of 
${\bf X}^s$ that are in region $\mathcal{R}_1$, i.e.,
\begin{equation*}
k_j = \left\{
\begin{array} {l l }
r_1-r_0{,} & \quad 1 \leq j \leq r_0 \\
r_1-j+1{,} & \quad r_0+1 \leq j \leq r_1 \\
0{,} & \quad r_1+1 \leq j \leq r_2 .\\
\end{array} \right.
\end{equation*}
Meanwhile, in the $j$-th column ($1 \leq j \leq r_2$) of ${\bf X}^s$, 
$m_j~= ~\max \{r_1+1, j\}$ and $r_2$ are the starting and ending layer 
indices of region $\mathcal{R}_2$, respectively.

In (\ref{eq:app_5}), for $1 \leq j \leq r_2$, i.e., the first $r_2$ 
columns of ${\bf X}^s$, we consider three cases {depending on the 
value of $k$}: 1)~$k=k_j$; 2)~$k_j+1 \leq k \leq \min \{r_2-r_0, 
r_2-j+1\}$; and 3)~$k \leq k_j-1$ or $k \geq \min \{r_2-r_0+1, 
r_2-j+2\}$.

For the first case, i.e., $k=k_j$, it can be seen that the event {that 
$s_j=k$} happens when the elements of the $j$-th column of ${\bf X}^s$ 
{that are in the region $\mathcal{R}_2$} are all zeros. Therefore, we 
have
\begin{equation}
\label{eq:app_6}
\Pr\{s_j = k\} = \prod_{m=m_j}^{r_2} (1-p_m).
\end{equation}

For the second case, i.e., $k_j+1 \leq k \leq \min \{r_2-r_0, r_2-j+1\}$, 
the event {that $s_j=k$} means that the $j$-th column of ${\bf X}^s$ has 
$(k-k_j)$ nonzero elements in {the} region $\mathcal{R}_2$. Denote indices 
of these $(k-k_j)$ nonzero elements as $a_1,a_2,\cdots,a_{k-k_j}$ with 
$a_1 < a_2 < \cdots < a_{k-k_j}$. So $a_1,a_2,\cdots,a_{k-k_j} \in 
\mathcal{A}_j\overset{\triangle}=\{m_j, m_j+1,\cdots,r_2\}$. We then have
{\small
\begin{align}
&{\Pr\{s_j = k\}} \nonumber \\
&= \sum_{\substack{a_1\!,a_2\!,\cdots\!,a_{k\!-\!k_j}\!\in\!\mathcal{A}_j \\ 
a_1\!<\!a_2\!<\!\cdots\!<\!a_{k\!-\!k_j}}} p_{a_1} \cdots p_{a_{k-k_j}} 
\!\prod_{\substack{m\!=\!m_j \\ m\!\neq\!a_i \\ i\!=\!1\!,\!\cdots\!,\!k\!
-\!k_j}}^{r_2}\!(1\!-\!p_m) \nonumber\\
&= \sum_{\substack{a_1\!,a_2\!,\cdots\!,a_{k\!-\!k_j}\!\in\!\mathcal{A}_j \\ 
a_1\!<\!a_2\!<\!\cdots\!<\!a_{k\!-\!k_j}}} {p_{a_1} \cdots p_{a_{k-k_j}} 
\over (1\!-\!p_{a_1}) \cdots (1\!-\!p_{a_{k\!-\!k_j}})}
\prod_{m\!=\!m_j}^{r_2} (1\!-\!p_m) \nonumber \\
&= \bigg[\!\prod_{m\!=\!m_j}^{r_2}\!(\!1\!-\!p_m\!)\!\bigg]
\sum_{\substack{a_1\!,a_2\!,\cdots\!,a_{\!k\!-\!k_j}\!\in\!\mathcal{A}_j \\ 
a_1\!<\!a_2\!<\!\cdots\!<\!a_{\!k\!-\!k_j}}}\!{p_{a_1}\!\cdots\!p_{a_{\!k\!
-\!k_j}} \over (\!1\!-\!p_{a_1}\!)\!\cdots\!(\!1\!-\!p_{a_{\!k\!-\!k_j}}\!)}.
\label{eq:app_8}
\end{align}
}

For the third case, i.e., $k \leq k_j-1$ or $k \geq \min \{r_2-r_0+1, 
r_2-j+2\}$, since $k_j \leq s_j \leq \min \{r_2-r_0, r_2-j+1\}$ for $1 
\leq j \leq r_2$, the event {that $s_j=k$} never happens, i.e.,
\begin{equation}
\label{eq:k_otherwise}
\Pr\{s_j = k\} = 0.
\end{equation}

According to (\ref{eq:app_3_1}) and (\ref{eq:app_5})-(\ref{eq:k_otherwise}) and
the fact that $l\ge r_1 - r_0 \ge k_j$, we have 
{\scriptsize
\begin{equation*}
\begin{array}{ll}
&\Pr\left\{\text{P is acceptable}\right\} \\
&\geq 1\!-\!\prod_{j=1}^{r_2} \bigg\{ \prod_{m=m_j}^{r_2} (1-p_m) \!+\!
\sum_{k\!=\!k_j\!+\!1}^{\min\{l, r_2-r_0, r_2-j+1\}}\!\bigg[\!\prod_{m\!
=\!m_j}^{r_2} (1\!-\!p_m)\!\bigg]\!\sum_{\substack{a_1\!,a_2\!,\cdots\!,
a_{k\!-\!k_j}\!\in\!\mathcal{A}_j \\ a_1\!<\!a_2\!<\!\cdots\!<\!a_{k\!-
\!k_j}}} {p_{a_1} \cdots p_{a_{k-k_j}} \over (1-p_{a_1}) \cdots 
(1-p_{a_{k-k_j}})}
\!\bigg\} \\
&= 1\!-\!\prod_{j\!=\!1}^{r_2}\!\bigg[\!\prod_{m\!=\!m_j}^{r_2} (1\!-
\!p_m) \!\bigg]  \cdot \prod_{j\!=\!1}^{r_2} \bigg\{\!1\!+\!
\sum_{k\!=\!k_j\!+\!1}^{\min\{l, r_2-r_0, r_2-j+1\}} \sum_{\substack{a_1\!,
a_2\!,\cdots\!,a_{k\!-\!k_j}\!\in\!\mathcal{A}_j \\ a_1\!<\!a_2\!<
\!\cdots\!<\!a_{k\!-\!k_j}}} {p_{a_1} \cdots p_{a_{k-k_j}} \over 
(1-p_{a_1}) \cdots (1-p_{a_{k-k_j}})}
\!\bigg\} \\
& \overset{\text{(a)}}= 1\!-\!\bigg[\!\prod_{j\!=\!1}^{r_1}\!\prod_{m\!
=\!r_1\!+\!1}^{r_2}\!(\!1\!-\!p_m\!)
\!\cdot\!\prod_{j\!=\!r_1\!+\!1}^{r_2}\!\prod_{m\!=\!j}^{r_2}\!(\!1\!
-\!p_m\!)\!\bigg] \cdot \prod_{j\!=\!1}^{r_2} \bigg\{\!1\!+\!
\sum_{k\!=\!k_j\!+\!1}^{\min\{l, r_2\!-\!r_0, r_2\!-\!j\!+\!1\}} 
\sum_{\substack{a_1\!,a_2\!,\cdots\!,a_{k\!-\!k_j}\!\in\!\mathcal{A}_j \\ 
a_1\!<\!a_2\!<\!\cdots\!<\!a_{k\!-\!k_j}}} {p_{a_1} \cdots p_{a_{k-k_j}} 
\over (1-p_{a_1}) \cdots (1-p_{a_{k-k_j}})}
\!\bigg\} \\
& = 1\!-\!\bigg[ \prod_{m=r_1+1}^{r_2} (1-p_m)^{r_1}
\cdot \prod_{m=r_1+1}^{r_2} (1-p_m)^{m-r_1} \bigg] \cdot 
\prod_{j\!=\!1}^{r_2} \bigg\{\!1\!+\!
\sum_{k\!=\!k_j\!+\!1}^{\min\{l\!, r_2\!-\!r_0\!,r_2\!-\!j\!+\!1\}} 
\sum_{\substack{a_1\!,a_2\!,\cdots\!,a_{k\!-\!k_j}\!\in\!\mathcal{A}_j \\ 
a_1\!<\!a_2\!<\!\cdots\!<\!a_{k\!-\!k_j}}} {p_{a_1}\!\cdots\!p_{a_{k\!-
\!k_j}} \over (\!1\!-\!p_{a_1}) \!\cdots\!(\!1\!-\!p_{a_{k\!-\!k_j}})}
\!\bigg\} \\
&=1\!-\!\bigg[ \prod_{m=r_1+1}^{r_2} (1-p_m)^m \bigg] \cdot 
\prod_{j\!=\!1}^{r_2} \bigg\{\!1\!+\!
\sum_{k\!=\!k_j\!+\!1}^{\min\{l, r_2-r_0, r_2-j+1\}} \sum_{\substack{a_1\!,
a_2\!,\cdots\!,a_{k\!-\!k_j}\!\in\!\mathcal{A}_j \\ a_1\!<\!a_2\!<\!\cdots
\!<\!a_{k\!-\!k_j}}} {p_{a_1}\!\cdots\!p_{a_{k\!-\!k_j}} \over (\!1\!-
\!p_{a_1}) \!\cdots\!(\!1\!-\!p_{a_{k\!-\!k_j}})}
\!\bigg\}.
\end{array}
\end{equation*}
}
where to obtain $\text{(a)}$ we have used the fact 
that $m_j = r_1+1$ for $1 \leq j \leq r_1$ and $m_j = j$ for $r_1+1 \leq j 
\leq r_2$. Using the facts that  $l=\lceil (r_0+r_2+1) / 2 \rceil$ and $p_m 
= e^{-\alpha(m-r_0-1)}$ for $r_1 +1 \leq m \leq r_2$, we obtain 
(\ref{e:propos_1_eq}).
{This completes the proof.}
\end{IEEEproof}

% Bibliography:
%\clearpage
\addcontentsline{toc}{chapter}{Bibliography}
\bibliographystyle{IEEEtran}
\bibliography{IEEEabrv,mybib,mypub}

\end{document}